\documentclass[prd,aps,twocolumn,showkeys,nofootinbib,superscriptaddress]{revtex4-2}

\usepackage{amsmath}
\usepackage{amsfonts}
\usepackage{amssymb}	
\usepackage{graphicx}
\usepackage{bm}
\usepackage{color}
\usepackage{ulem}
\usepackage{commath}
\usepackage[T1]{fontenc}
\usepackage{lmodern}
\usepackage{xcolor}
\usepackage{hyperref}
\allowdisplaybreaks
\usepackage{aas_macros} 
\usepackage{orcidlink}

\def\TEOBResumS{\texttt{TEOBResumS}}
\def\ph{\varphi}
\def\p{\partial}
\def\eob{{\rm EOB}}

\def\rmd{{\rm d}}

\newcommand\be{\begin{equation}}
\newcommand\ee{\end{equation}}

\newcommand{\gr}[1]{{\textcolor{gray}{#1} }}

\begin{document}

\title{Comparing effective-one-body and Mathisson-Papapetrou-Dixon results for a spinning test particle on circular equatorial orbits around a Kerr black hole}

\author{Angelica \surname{Albertini}\,\orcidlink{0000-0002-9556-1323}}
\affiliation{Astronomical Institute of the Czech Academy of Sciences,
Bo\v{c}n\'{i} II 1401/1a, CZ-141 00 Prague, Czech Republic}
\affiliation{%
Institute of Theoretical Physics
Faculty of Mathematics and Physics, Charles University, 
V Holešovičkách 2,
180 00, Prague, Czech Republic 
}

\author{Viktor \surname{Skoup\'{y}}\,\orcidlink{0000-0001-7475-5324}}
\affiliation{%
Institute of Theoretical Physics
Faculty of Mathematics and Physics, Charles University, 
V Holešovičkách 2,
180 00, Praha 8, Czech Republic 
}

\author{Georgios \surname{Lukes-Gerakopoulos}\,\orcidlink{0000-0002-6333-3094}}
\affiliation{Astronomical Institute of the Czech Academy of Sciences,
Bo\v{c}n\'{i} II 1401/1a, CZ-141 00 Prague, Czech Republic}

\author{Alessandro \surname{Nagar}\,\orcidlink{0000-0001-7998-2673}}
\affiliation{INFN Sezione di Torino, Via P. Giuria 1, 10125 Torino, Italy} 
\affiliation{Institut des Hautes Etudes Scientifiques, 91440 Bures-sur-Yvette, France}

\begin{abstract}
We consider a spinning test particle around a rotating black hole and compare the Mathisson-Papapetrou-Dixon (MPD) formalism under the Tulczyjew-Dixon spin supplementary condition to the test-mass limit of the effective-one-body (EOB) Hamiltonian of [Phys.~Rev.~D.90,~044018(2014)], with enhanced spin-orbit sector. 
We focus on circular equatorial orbits: we first compare the constants of motion at their linear in secondary spin $\sigma$ approximation and then we compute the gravitational-wave (GW) fluxes using a frequency domain Teukolsky equation solver. We find no difference between the EOB and MPD fluxes when the background spacetime is Schwarzschild, while the difference for a Kerr background is maximum for large, positive spins. Our work could be considered as a first step to improve the radiation reaction of the EOB model, in view of the needs of the next-generation of GW detectors.
\end{abstract}

\date{\today}

\maketitle

\section{Introduction}

Since the first detection of gravitational waves (GWs) in 2015 \cite{Abbott:2016nmj}, the LIGO-Virgo-KAGRA collaboration has been mainly observing GWs coming from compact binary coalescences, where the binaries' mass ratio is roughly of the order $\mathcal{O}(1)$, i.e. comparable-mass binaries. The upcoming generation of GW observatories, either ground- or space-based, will instead reach lower frequencies and allow us to observe a more diverse collection of sources. Among these are in particular intermediate- and extreme-mass-ratio inspirals (IMRIs and EMRIs respectively), systems in which a stellar-mass compact object is revolving around a massive or super massive black hole. These systems are expected to have a richer phenomenology, and their detection and parameter estimation will, thus, require additional effort on the GW waveform modeling side.

Among the currently used models for comparable-mass binaries are effective-one-body (EOB) ones: these are based on a Hamiltonian conservative dynamics complemented by an analytical radiation reaction force. Both for the conservative and for the dissipative sector these models make use of post-Newtonian (PN) expressions that are resummed in various ways for better convergence in the strong field. In order to choose a proper resummation, the functions are usually compared to test-mass results. For instance, in Ref.~\cite{Nagar:2019wrt} the authors proposed an efficient resummation scheme that was chosen to match numerical results for a spinning particle moving in a Schwarzschild black hole background.

The present work is a preliminary step to such a resummation scheme within an EOB model. It aims at analyzing the dynamics of a spinning particle on circular equatorial orbits (CEOs) around a Kerr black hole and at evaluating the related gravitational wave fluxes. A similar analysis has already been carried out
within the Mathisson-Papapetrou-Dixon (MPD) formalism under different spin supplementary conditions (SSCs) \cite{Lukes-Gerakopoulos:2017vkj} and in comparison with PN results \cite{Harms:2015ixa}. Actually, Ref.~\cite{Harms:2016ctx} provided a precursor study to the one we are presenting here, but constrained to a spinning particle in a Schwarzschild black hole background, and it constitutes the first step of what was carried out in the aforementioned Ref.~\cite{Nagar:2019wrt}. Instead of taking into account three different SSCs as done in Ref.~\cite{Harms:2016ctx}, we only consider the Tulczyjew-Dixon one \cite{tulczyjew1959motion,Dixon:1970zza}, and we employ a frequency-domain (FD) Teukolsky equation (TE) solver instead of the time-domain one used in Ref.~\cite{Harms:2016ctx}. 

This paper is organized as follows. In Sec.~\ref{sec:EOBframe} we give an overview of the EOB formalism, describing its test-particle limit and evaluating the relation between the EOB radius and the orbital frequency for CEOs at liner order in the secondary spin $\sigma$. In Sec.~\ref{sec:MPD} we present the MPD formalism, the conditions for CEOs, and the analogous radius-frequency relation as computed for EOB. In Sec.~\ref{sec:consts} we compare the constants of motion, presenting numerical results and also computing the PN expansion of the angular momentum for both approaches. In Sec.~\ref{sec:LSO} the behavior of the last stable orbit (LSO) for EOB and MPD is compared. In Sec.~\ref{sec:fluxes} we finally present the fluxes evaluated with the FD TE solver. In Sec.~\ref{sec:conclusions} we gather our concluding remarks. In Appendix~\ref{sec:PN_HF} we discuss the behavior of the horizon flux exploiting an analytical PN formula, and in Appendix~\ref{sec:Schwarzschild} we discuss the differences between this work and Ref.~\cite{Harms:2016ctx}.

{\bf Notation and conventions:} We define the mass ratio $q \equiv m_1/m_2 \ge 1$,  the symmetric mass ratio $\nu\equiv m_1 m_2/M^2$, where $m_{1,2}$ are the masses of the two bodies, $M\equiv m_1+m_2$ and we use the convention $m_1\geq m_2$. We address with $(S_1,S_2)$ the individual, dimensionful, spin components aligned along the direction of the orbital angular momentum. In the test-particle limit $m_1 \gg m_2$, $M \sim m_1$ is the mass of the central black hole, and $\mu = m_2$ is the mass of the particle. The dimensionless spin of the central black hole is then defined as $\hat{a} \equiv S_1/M^2$, and the dimensionless spin of the particle is $\sigma \equiv S_2/(\mu M)$. We use geometric units with $G=c=1$, normalize quantities by the total mass $M$ (e.g., the time is $t\equiv T/M$, the radial separation is $r\equiv R/M$ etc.), and in practice work numerically with $\mu = M = 1$.

\section{The effective-one-body framework} \label{sec:EOBframe}

The effective-one-body approach~\cite{Buonanno:1998gg,Buonanno:2000ef,Damour:2000we,Damour:2001tu,Damour:2015isa} to the two-body problem in general relativity is based on a map between the motion of two compact objects into the motion of a single particle in an effective metric. This mapping yields the Hamiltonian of the system, which encodes the conservative part of the dynamic, and reads
\be
\label{eq:HEOB}
\hat{H}_\eob = \frac{H_\eob}{\mu} = \frac{1}{\nu}\sqrt{1+2\nu(\hat{H}_{\rm eff} -1)},
\ee
where $\hat{H}_{\rm eff} \equiv H_{\rm eff} / \mu$ is the effective Hamiltonian. 
For a full evolution, one has to take into account the energy and angular momentum carried away by gravitational waves to infinity and, if the compact bodies are black holes, into their horizons. This dissipative contribution to the evolution in the EOB formalism is taken into account via an \textit{analytical} expression for the radiation reaction $\hat{\mathcal{F}}_\ph$, which enters the right-hand side of the Hamiltonian equations of motion:
\begin{subequations}
\begin{align}
\dot{\ph} &= \Omega = \p_{p_\ph} \hat{H}_\eob, \\
\dot{r} &= \left( \frac{A}{B} \right)^{1/2} \p_{p_{r_*}} \hat{H}_\eob, \\
\dot{p}_\ph &= \hat{\mathcal{F}}_\ph , \\
\dot{p}_{r_*} &= - \left( \frac{A}{B} \right)^{1/2} \p_{r} \hat{H}_\eob .
\end{align}
\end{subequations}
By solving these equations one finds the evolution of the dynamical variables, that allows finally to evaluate the gravitational waveform. 

Since we are focusing on the motion of a spinning particle around a Kerr black hole, we are interested in the EOB Hamiltonian for spinning black hole binaries.
The first EOB Hamiltonian for such binaries was presented in Ref.~\cite{Damour:2001tu}. The spin-orbit coupling was then extended to next-to-leading order in Ref.~\cite{Damour:2007nc}, and to next-to-next-to-leading order in Ref.~\cite{Nagar:2011fx}. A complete EOB model for coalescing spinning binaries, including the radiation reaction, was finally presented in Ref.~\cite{Damour:2014sva}.
On the latter is based the EOB model for spinning black hole binaries that will be the basis of our analysis, \TEOBResumS{}~\cite{Nagar:2018zoe, Nagar:2020pcj}.
In this work we restrict our analysis to the test-mass limit of the Hamiltonian of Ref.~\cite{Damour:2014sva}, with few modifications that will be described in the following. We would like to highlight here that in this lineage of papers the Hamiltonian is derived by combining the Arnowitt-Deser-Misner (ADM) Hamiltonian formalism for the orbital degrees of freedom with a non-covariant treatment of the spin degrees of freedom, and that so far the spin-orbit couplings in the test-particle limit are reproduced only through 2.5PN order, as pointed out in Ref.~\cite{Buonanno:2009qa}. There is as well a different branch of EOB spinning Hamiltonians, stemming from Refs.~\cite{Barausse:2009aa,Barausse:2009xi, Barausse:2011ys}, that in the test-particle limit reproduces all spin-orbit couplings and uses the Newton-Wigner SSC \cite{Newton:1949RvMP...21..400N}. 

\subsection{Test-particle limit}
\label{sec:TPlim}

The test-particle limit of the EOB Hamiltonian of Ref.~\cite{Damour:2014sva} was previously presented in Sec.~III of Ref.~\cite{Harms:2016ctx}, that was focussing on the dynamics of a spinning particle around a Schwarzschild black hole. The Hamiltonian used in this work will be the one presented in Eq.~(70) therein, but we rederive it here for completeness. We start from the \TEOBResumS{} Hamiltonian defined for comparable-mass black hole binaries with aligned spins. For a test particle, the EOB Hamiltonian in Eq.~\eqref{eq:HEOB} simply reduces to the effective Hamiltonian, that for circular orbits reads
\begin{align}
\hat{H}_{\rm eff} &= \hat{H}_{\rm eff}^{\rm orb} +  \hat{H}_{\rm eff}^{\rm SO}, \\ 
\hat{H}_{\rm eff}^{\rm orb} &= \sqrt{A (1 + p_\ph^2 u_c^2)}  , \\
\hat{H}_{\rm eff}^{\rm SO} &= p_\ph \left(G_S \hat{S} + G_{S_*} \hat{S}_* \right) ,
\end{align}
where the spin variables are defined as
\begin{align}
    \hat{S} &= \frac{S_1}{M^2} + \frac{S_2}{M^2} , \\
    \hat{S}_* &= \frac{m_2}{m_1} \frac{S1}{M^2} +  \frac{m_1}{m_2}  \frac{S_2}{M^2} ,
\end{align}
and we note that the second definition is equivalent to $\hat{S}_* = \nu [S_1/(m_1^2) + S_2 / (m_2^2)]$. Here $p_\varphi$ is the orbital angular momentum, $G_S, G_{S_*}$ are the gyro-gravitomagnetic functions, and $A$ is the main EOB radial potential, defined as 
\be
A = \frac{1 - 4 u_c^2}{1 + 2 u} ,
\ee
where $u_c\equiv1/r_c$ is the inverse centrifugal radius. Differently from Ref.~\cite{Damour:2014sva}, in this work we also take into account the next-to-next-to-leading-order (NNLO) contribution to $r_c$ as evaluated in Ref.~\cite{Nagar:2018plt}, so that the centrifugal radius reads
\be
r_c = \left(r^2 + (\hat{a}+\sigma)^2 \left(1+\frac{2}{r} + \frac{\delta a^2_{\rm NLO}}{r} + \frac{\delta a^2_{\rm NNLO}}{r^2}\right) \right)^{1/2} .
\ee
The gyrogravitomagnetic functions are 
\begin{align}
G_S &= 2 u u_c^2 , \\
G_{S_*} &= \left( \frac{1}{r_c^2} \right) \left[ \frac{r_c \nabla \sqrt{A_{\rm eq}} }{1 + \sqrt{Q}} + \frac{(1 - \nabla r_c) \sqrt{A_{\rm eq}}}{\sqrt{Q}}  \right] , \label{eq:GSs}
\end{align}
with
\begin{align}
\nabla &\equiv (B_{\rm eq})^{-1/2} d/dr \\
Q &= 1+ p_\varphi^2 u_c^2 \\
B_{\rm eq} &= \frac{u_c^2}{u A} .
\end{align}
For $G_{S_*}$ we adopt the complete zeroth-order GSF contribution, determined by the spin-orbit coupling of a spinning test-particle in a Kerr background, as appears in Eq.~(2.21) in Ref.~\cite{Bini:2015xua}. This expression differs from the test-mass limit of the $G_{S_*}$ used in the standard comparable-mass version of \TEOBResumS{}, that is just the leading order of the PN expansion of the complete one, corresponding to $(3/2) u^3$. This choice in the standard \TEOBResumS{} is due to the use of the so-called Damour-Jaranowski-Sch\"afer (DJS) spin gauge\footnote{We remind that the necessity of choosing a spin gauge in EOB is connected to the choice of a center of mass of an extended body in a curved spacetime. This is clearly seen in Ref.~\cite{Damour:2008qf}, where to obtain the EOB (real) Hamiltonian for spinning binaries (with NLO spin-orbit coupling), one starts from the related ADM Hamiltonian in the center-of-mass frame and performs two canonical transformations. The first one is needed to go to EOB coordinates, while the second one only affects NLO spin-orbit terms, and involves two arbitrary $\nu$-dependent dimensionless coefficients. These coefficients can be viewed as gauge parameters, connected to the arbitrariness of defining a local frame to measure the spin vectors. The real EOB Hamiltonian is then related to the effective one via the usual energy map.}~\cite{Damour:2000we, Damour:2008qf}, which requires the gyrogravitomagnetic functions not to depend on $p_\varphi$. In Ref.~\cite{Albertini:2024rrs}, we explored the performance of using another spin gauge, referred to as anti-DJS. In this gauge, the expression in Eq.~\eqref{eq:GSs} is used as a prefactor in the PN expansion of $G_{S_*}$, so that the Hamiltonian we use here is actually the test-mass limit of the one used in Ref.~\cite{Albertini:2024rrs}.

However, if one took directly the $\nu \rightarrow 0$ limit, due to the way $\hat{S}_*$ is defined, the second spin-orbit coupling function $G_{S_*}$ would not contribute to the Hamiltonian. Following Ref.~\cite{Harms:2016ctx}, we consider instead that when $m_1 \gg m_2$ and $M \sim m_1$, the spin-orbit effective Hamiltonian in terms of $m_1,m_2$ becomes
\begin{align}
\label{eq:HSOeff}
\hat{H}_{\rm eff}^{\rm SO} &= p_\varphi \left[  G_S \left( \frac{S_1}{m_1^2} + \frac{S_2}{m_1^2} \frac{m_2}{m_2} \right) \right.\nonumber \\
                           &+ \left. G_{S_*} \left( \frac{m_2}{m_1} \frac{S1}{m_1^2} +  \frac{m_1}{m_2}  \frac{S_2}{m_1^2} \right) \right]\nonumber \\
                           &= p_\varphi \left[  G_S \left( \hat{a} +  \frac{m_2}{m_1} \sigma \right)+ G_{S_*} \left( \frac{m_2}{m_1} \hat{a} +  \sigma \right) \right]
\end{align}
where $\hat{a} \equiv a/M = S_1/(M^2) \sim S_1/(m_1^2)$. At this point of the calculation, we can safely take the test-mass limit $(m_2/m_1\rightarrow 0)$ of all the quantities in the last line in Eq.~\eqref{eq:HSOeff} and thus recover
\be
\hat{H}_{\rm eff}^{\rm SO} = p_\varphi \left(  G_S \hat{a} + G_{S_*} \sigma \right).
\ee

\subsection{Relation between radius and frequency for CEOs} \label{sec:RadFr}

One of our aims in the following will be to compare numerical gravitational wave fluxes evaluated with a frequency domain (FD) Teukolsky equation (TE) solver, that takes as input the energy and the angular momentum for CEOs. For this purpose, we need analytical expression for these dynamical quantities within the EOB and MPD formalisms at linear order in the spin of the particle. These expressions are analytically found as functions of the inverse radial coordinate $u$. However, the EOB radial coordinate only coincides with the Boyer-Lindquist one if $\sigma=0$, namely for a nonspinning particle. We thus need to express all the quantities as functions of a gauge-invariant parameter. We choose to exploit the frequency parameter $x = \Omega^{2/3}$, where $\Omega = \dot{\ph}$ is the orbital frequency, since the relation $u(x)$ can be easily derived for CEOs at linear order in $\sigma$. This will allow us to finally express the energy and the angular momentum as functions of $x$.

If the central black hole is nonspinning ($\hat{a}=0$), the Kepler law holds for circular orbits and $x=u$, so that $u_{\rm EOB} = u_{\rm MPD}$. For a nonspinning particle on a Kerr black hole, the relation between the inverse radial coordinate and the frequency parameter both for MPD and EOB reads
\be
u(x)=\frac{x}{(1- \hat{a} x^{3/2})^{2/3}}
\ee
so that again the radial coordinates coincide. For a spinning particle on a Schwarzchild black hole, the relationships $u_{\rm EOB, MPD} (x)$ have been derived in Ref.~\cite{Harms:2016ctx}, and read
\begin{align}
  u_{\rm EOB} (x) &= x + \frac{x^{5/2}}{1 + \sqrt{\frac{1 - 2x}{1-3x}}} \sigma  \\
  u_{\rm MPD} (x) &= x + x^{5/2} \sigma
\end{align}

The purpose of the current section is to derive the relation between the frequency parameter $x$ and the EOB inverse radial coordinate $u_{\rm EOB}$ for a spinning particle on CEOs around a Kerr black hole. For simplicity, throughout this section we will simply write $u$ instead of $u_{\rm EOB}$. We start by imposing the circular condition
\be
\label{eq:EOB_COC}
\frac{\p \hat{H}_{\rm eff}}{\p u} = 0
\ee
and substitute in this expression an ansatz for the angular momentum to be linear in $\sigma$, namely $p_{\ph} \rightarrow l_0 + \sigma l_1$. Then we solve for $l_1$. Here, $l_0$ is the angular momentum for a non-spinning test particle on a Kerr black hole, that reads
\be
l_0 = \frac{1 - 2 \hat{a} u^{3/2} + a^2 u^2}{\sqrt{u (1 - 3u + 2 \hat{a} u^{3/2} )}}.
\ee

The orbital frequency is found as
\be
\Omega = \frac{\p \hat{H}_{\rm eff}}{\p p_\ph} .
\ee
Substituting $p_{\ph} \rightarrow l_0 + \sigma l_1$ and expanding in $\sigma$, this gives $\Omega  = \Omega_{\rm I} + \sigma \Omega_{\rm II}$ with
\begin{align}
\Omega_{\rm I} &= \frac{l_0 u_c^2 A}{\sqrt{A (1 + l_0^2 u_c^2)}} + \hat{a} G_S \label{eq:OmgI}\\
\Omega_{\rm II} &= \frac{l_1 u_c^2 A^2}{[A (1 + l_0^2 u_c^2)]^{3/2}} + \nonumber \\
         &+ G_{S_*} |_{p_\ph = l_0} + l_0 \left. \frac{\p G_{S_*}}{\p p_\ph} \right|_{p_\ph = l_0} . \label{eq:OmgII}
\end{align}
The quantities in $\Omega_{\rm II}$ are all evaluated at zeroth order in $\sigma$, while $\Omega_{\rm I}$ can further be expanded considering that $u_c$, $A$ and $G_S$ also contain $\sigma$. This leads us to obtain $\Omega =\Omega_0 + \sigma \Omega_1$, where $\Omega_0$ is Eq.~\eqref{eq:OmgI} evaluated with the zeroth-order-in-$\sigma$ contributions $u_{c,0}$, $A_0$ and $G_{S,0}$, while $\Omega_1$ includes both Eq.~\eqref{eq:OmgII} and the linear-in-$\sigma$ contribution to Eq.~\eqref{eq:OmgI}.

The frequency parameter $x$ can be formally expanded as
\begin{align}
\label{eq:Omg01}
x &= (\Omega_0 + \sigma \Omega_1)^{2/3} \nonumber \\ 
&\simeq \Omega_0^{2/3} + \frac{2}{3}\sigma \frac{\Omega_1}{\Omega_0^{1/3}} + O(\sigma^2).
\end{align}
To get $u_{\rm EOB}(x)$ we can consider then that at linear order in $\sigma$
\begin{align}
\label{eq:u01}
u(x) &= u_0(x) + \sigma u_1(x) \nonumber \\
     &= u_0(x) + \sigma \left(-\frac{\p x}{\p \sigma} \frac{\p u_0}{\p x}\right) ,
\end{align}
where $\p x / \p \sigma = 2\Omega_1/(3 \Omega_0^{1/3}) $. For practical reasons we do not present the final formula here, and choose to include it in the supplementary material attached to this work~\cite{supp}.

\section{MPD} \label{sec:MPD}

In this section we provide the key features of the MPD formalism, for more see Refs.~\cite{Semerak:1999MNRAS.308..863S,Costa18}. The MPD formalism \cite{Mathisson:1937zz, Papapetrou:1951pa, Dixon:1964cjb} provides a framework to describe the motion of an extended test body in a curved background. This extended body can be reduced to the pole-dipole approximation, in which we ignore all the higher multipoles of the body. In the pole-dipole approximation, when only the gravitational interaction is taken into account, the MPD equations read
\begin{align} 
 \label{eq:MPD}
 \frac{{\rm D}P^\mu}{\rmd \tau} &= 
 - \dfrac{1}{2} \; {R^\mu}_{\nu\rho\sigma} \; v^\nu \; S^{\rho\sigma} \; , \nonumber \\ 
 \frac{{\rm D}S^{\mu\nu}}{\rmd \tau} & = 
 P^\mu v^\nu - P^\nu v^\mu \; ,
\end{align}
where $P^\mu$ is the four-momentum of the particle, $R^\mu{}_{\nu\rho\sigma}$ is the Riemann tensor of the background spacetime, $v^\mu = \rmd x^\mu/\rmd \tau$ is the four-velocity, $S^{\mu\nu}$ is the spin tensor of the particle and ${\rm D}/\rmd \tau = v^\mu \nabla_\mu$ is the covariant derivative along the worldline parametrized by the proper time $\tau$.

The system of equations \eqref{eq:MPD} is underdetermined and we need three extra constraints to close it. Actually, if he had not chosen the proper time as our affine parameter, we would need four constraints, but the proper time choice implies the $v^\mu v_\mu=-1$ four-velocity constraint. The three remaining constraints fix the centre of mass of the extended body, which is used as a reference point in the extended body with respect to which the multipoles of the body are calculated. This point is called centroid and is specified by a SSC in the form $S^{\mu\nu} V_\mu = 0$, where $V_\mu$ is a timelike vector field. In this work, we use the Tulczyjew-Dixon (TD) SSC \cite{tulczyjew1959motion,Dixon:1970zza} 
\begin{align}
 S^{\mu\nu} P_\mu = 0  \; .
\end{align}
Since the extended test body is represented by a point, it is often called a particle, and in particular in the pole-dipole case a spinning particle.  

Having chosen the TD centroid, allows us to define the spin four-vector
\begin{align}
\label{eq:SpinVect}
 S_\mu = -\frac{1}{2} \epsilon_{\mu\nu\rho\sigma}
          \, u^\nu \, S^{\rho\sigma} \; ,
\end{align}
where $\epsilon_{\mu\nu\rho\sigma}$ is the Levi-Civita tensor and $u^\nu := P^\nu/m_p$ is the specific four-momentum. The inverse relation of this equation reads 
\begin{align}
   S^{\rho\sigma}=-\epsilon^{\rho\sigma\gamma\delta} S_{\gamma} u_\delta \; . 
   \label{eq:SpinTens}
\end{align}
The spin four-vector allows us to understand better the orientation of the spin as we shall see later on. 

Under the TD SSC, the mass $m_p$ of the particle with respect to the four-momentum
\begin{equation} \label{eq:p_norm}
    {m_p}^2 = -P^\mu P_\mu
\end{equation}
and the magnitude of the spin 
\begin{align}
 \label{eq:spinMagnitude}
 S^2 = \frac{1}{2} S^{\mu\nu}S_{\mu\nu}=S^\mu S_\mu
\end{align}
are conserved quantities (see, e.g., \citep{Semerak:1999MNRAS.308..863S}). The conservation of the above quantities is independent of the spacetime background. On the other hand, the symmetries of the spacetime introduce for each Killing vector $\xi^\mu$ a specific quantity 
\begin{align}\label{eq:MPD_killsym}
 C=\xi^\mu P_\mu-\frac12 \xi_{\mu;\nu} S^{\mu\nu} \, ,
\end{align}
which is conserved upon the evolution of the MPD equations.

Even if we have chosen a SSC, it is not so obvious how to evolve the MPD equations, since there is no evolution equation for the four-velocity. For the TD SSC, however, exists a relation of the four-velocity in terms of the four-momentum and the spin tensor \cite{Ehlers1977}, which reads 
\begin{align}
 \label{eq:v_p_TUL}
v^\mu = \frac{\textsf{m}_v}{m_p} \left(
           u^\mu + 
           \frac{ 2 \; S^{\mu\nu} R_{\nu\rho\kappa\lambda} u^\rho S^{\kappa\lambda}}
           {4 {m_p}^2 + R_{\alpha\beta\gamma\delta} S^{\alpha\beta} S^{\gamma\delta} }
           \right)  \; ,
\end{align}
where $m_v = -P_\mu v^\mu$ is the rest mass with respect to the four-velocity $v^\mu$. This mass  $m_v$ is not conserved under the TD SSC an it can be used so that the constraint $v^\mu v_\mu = -1$ during the MPD evolution is satisfied leading to 
\begin{equation} \label{eq:mass_vmu}
    m_v = \frac{\mathcal{A} {m_p}^2}{\sqrt{\mathcal{A}^2 {m_p}^2 - \mathcal{B}S^2}} \; ,
\end{equation}
\cite{Witzany:2019}, where
\begin{align}
     \mathcal{A} &= 4 {m_p}^2 + R_{\alpha\beta\gamma\delta} S^{\alpha\beta} S^{\gamma\delta} \; , \\
     \mathcal{B} &= 4 h^{\kappa\eta} R_{\kappa\iota\lambda\mu} P^\iota S^{\lambda\mu} R_{\eta \nu \omega \pi} P^\nu S^{\omega\pi} \; , \\
     h^\kappa{}_\eta &= \frac{1}{S^2} S^{\kappa\rho} S_{\eta\rho} \; .
 \end{align}

\subsection{The Kerr spacetime background}

Since our work deals with the motion of a spinning in the Kerr spacetime, let us briefly introduce this spacetime. The Kerr geometry in BL coordinates $(t,r,\theta,\phi)$ is described by the metric
\begin{multline}
    \rmd s^2 = g_{tt}~\rmd t^2+2~g_{t\phi}~\rmd t~\rmd \phi + g_{\phi\phi}~\rmd \phi^2 \\
       + g_{rr}~\rmd r^2+g_{\theta\theta}~\rmd \theta^2 \; , \label{eq:LinEl}
\end{multline}
where the metric coefficients are
 \begin{eqnarray}
   g_{tt} &=&-\left(1-\frac{2 M r}{\Sigma}\right) \; ,\nonumber\\ 
   g_{t\phi} &=& -\frac{2 a M r \sin^2{\theta}}{\Sigma} \; ,\nonumber\\
   g_{\phi\phi} &=& \frac{(\varpi^4-a^2\Delta \sin^2\theta) \sin^2{\theta}}{\Sigma} \; , \label{eq:KerrMetric}\\
   g_{rr} &=& \frac{\Sigma}{\Delta} \; ,\nonumber\\
   g_{\theta\theta} &=& \Sigma \nonumber
 \end{eqnarray} 
with
 \begin{eqnarray}
  \Sigma &=& r^2+ a^2 \cos^2{\theta} \; ,\nonumber\\
  \Delta &=& \varpi^2-2 M r \; ,\nonumber \\ 
  \varpi^2 &=& r^2+a^2 \; . \label{eq:Kerrfunc} 
 \end{eqnarray}

The Kerr spacetime is stationary and axisymmetric. This provides two Killing vector fields, the timelike one $\xi^\mu_{(t)}$ and the spacelike one $\xi^\mu_{(\phi)}$. Due to these Killing vector fields, Eq.~\eqref{eq:MPD_killsym} provides two constants of motion. In particular, thanks to the timelike field, the energy
 \begin{align}\label{eq:EnCons}
  E &= -P_t+\frac12g_{t\mu,\nu}S^{\mu\nu}
 \end{align}
is conserved, and thanks to the spacelike field, the component of the total angular momentum parallel to the rotational axis of Kerr ($z$ axis)
 \begin{align}\label{eq:JzCons}
  J_z &= P_\phi-\frac12g_{\phi\mu,\nu}S^{\mu\nu}
 \end{align}
is conserved. These two conserved quantities can be used to parametrize the spinning particles orbits.

The stress-energy tensor $T^{\mu\nu}$ of a MPD pole-dipole body
reads \citep{Faye:2006gx}
 \begin{equation}
     T^{\mu\nu} =  \frac{P^{(\mu} v^{\nu)}}{v^t} \frac{\delta^3}{\sqrt{-g}} - \nabla_\alpha \left( \frac{S^{\alpha(\mu} v^{\nu)}}{v^t} \frac{\delta^3}{\sqrt{-g}} \right)\, ,
\end{equation}
 where for Boyer-Lindquist (BL) coordinates $\delta^3 = \delta(r-r_p(t)) \delta(\theta-\theta_p(t)) \delta(\phi-\phi_p(t))$ is the delta function located at the particle's position.

\subsection{CEOs}

This work focuses on CEOs. For the MPD equations under TD SSC, to restrict the motion on the equatorial plane $\theta=\pi/2$, i.e. to ensure that $P^\theta=v^\theta=0$,
\begin{equation}
    S_\mu= - r S \,\delta_\mu^\theta \,,
\end{equation}
has to hold \cite{Skoupy:2021asz}, while the circularity of the orbits is ensured by $p^r=v^r=0$ \cite{Tod76}. 

For CEOs it holds that
\begin{align}
    E&=\frac{(a M S-m_p r^3)p_t+M S p_\phi}{m_p r^3} \label{eq:CEOE}, \\
    J_z&=\frac{S (M a^2-r^3)p_t+(m_p r^3+a M S)p_\phi}{m_p r^3} \label{eq:CEOJz}.
\end{align}

\subsection{Relationship between radius and frequency}

The relation $u_{\rm MPD}(x)$ for a spinning particle on a Kerr black hole has not been presented before in the literature, so we report it here. To obtain it, we exploit the expression for the orbital frequency $\Omega$ shown in Eq.~(22) of Ref.~\cite{Harms:2015ixa}, expand it at linear order in $\sigma$ and then consider Eqs.~(\ref{eq:Omg01},~\ref{eq:u01}). We find
\begin{align}
\label{eq:uMPDofx}
u_{\rm MPD}(x) &= \frac{1}{(1- \hat{a} x^{3/2})^2} \left[ x (1- \hat{a} x^{3/2})^{1/3} - \hat{a}\sigma x^3 \right . \nonumber \\
& \left . - (\hat{a} - \sigma) (1- \hat{a} x^{3/2})^{1/3} x^{5/2} \right] . 
\end{align}

\section{Constants of motion}
\label{sec:consts}

\begin{table*}[htp!]
\begin{center} 
\begin{ruledtabular}
\begin{tabular}{c c c | c c | c c}
$\hat{a}$ & $\sigma$ & $x$ & $J_z^{\rm MPD}$ & $p_\varphi^{\rm EOB} + \sigma$ & $E^{\rm MPD}$ & $E^{\rm EOB}$\\
\hline 
\hline
$-0.9$ & $-0.9$ & 0.05 & 4.26775 & 4.26804 & 0.977694 & 0.977696 \\
& & 0.1 & 3.61933 & 3.62097 & 0.966711 & 0.966743 \\
& & 1/6 & \gr{4.1848} & \gr{4.1908} & \gr{0.998211} & \gr{0.998472} \\
& & 0.2 & \gr{5.03648} & \gr{5.05246} & \gr{1.06611} & \gr{1.06719} \\
& $0.$ & 0.05 & 5.02994 & 5.02994 & 0.977042 & 0.977042 \\
& & 0.1 & 4.20851 & 4.20851 & 0.962393 & 0.962393 \\
& & 1/6 & \gr{4.46614} & \gr{4.46614} & \gr{0.978429} & \gr{0.978429} \\
& & 0.2 & \gr{5.09833} & \gr{5.09833} & \gr{1.02896} & \gr{1.02896} \\
& $0.9$ & 0.05 & 5.79212 & 5.79184 & 0.97639 & 0.976388 \\
& & 0.1 & 4.79769 & 4.79604 & 0.958075 & 0.958043 \\
& & 1/6 & 4.74748 & 4.74147 & 0.958648 & 0.958386 \\
& & 0.2 & \gr{5.16019} & \gr{5.14421} & \gr{0.991806} & \gr{0.990733} \\
\hline
$0.9$ & $-0.9$ & 0.05 & 3.89305 & 3.89382 & 0.975896 & 0.975902 \\
& & 0.1 & 2.72548 & 2.73173 & 0.953804 & 0.953935 \\
& & 1/6 & 2.17889 & 2.2044 & 0.928607 & 0.929723 \\
& & 0.2 & 2.05008 & 2.08796 & 0.918596 & 0.920683 \\
& $0.$ & 0.05 & 4.69301 & 4.69301 & 0.975458 & 0.975458 \\
& & 0.1 & 3.42877 & 3.42877 & 0.951351 & 0.951351 \\
& & 1/6 & 2.74778 & 2.74778 & 0.919528 & 0.919528 \\
& & 0.2 & 2.54676 & 2.54676 & 0.903835 & 0.903835 \\
& $0.9$ & 0.05 & 5.49297 & 5.4922 & 0.97502 & 0.975014 \\
& & 0.1 & 4.13205 & 4.12581 & 0.948898 & 0.948766 \\
& & 1/6 & 3.31666 & 3.29115 & 0.910449 & 0.909333 \\
& & 0.2 & 3.04344 & 3.00556 & 0.889075 & 0.886988 \\
\end{tabular}
\end{ruledtabular}
\end{center} 
\caption{Values of the energy and the angular momentum for circular orbits for fixed values of the frequency parameter $x$ (third column), both for EOB (fourth and sixth columns) and for MPD (fifth and seventh columns). For negative $\hat{a}$, the gray values are beyond LSO. For nonspinning secondary, as expected, the EOB and MPD values coincide. On the other hand, for positive values of $\sigma$, the EOB values are always smaller than the MPD ones, while for negative values of $\sigma$ the inverse is true. In both cases, the difference is larger when $\hat{a}$ is positive.}
\label{tab:consts}
\end{table*}

Having obtained the $u(x)$ relationship both for EOB and MPD, it is now straightforward to evaluate the energy and angular momentum in both frameworks at linear order in $\sigma$. Before discussing the numerical values, we note that we will compare Eq.~\eqref{eq:CEOJz} for the MPD total angular momentum to the EOB orbital angular momentum $p_\varphi$ (found before as $l_0 + \sigma l_1$) summed to $\sigma$. To explain our choice, we can look at the 3PN expansions of both quantities, that read

\begin{widetext}
\begin{align}
J_z^{\rm MPD} &= \frac{1}{\sqrt{x}}+\sigma +\frac{3 \sqrt{x}}{2}-\frac{5}{6} x (4 \hat{a}+3 \sigma )+x^{3/2} \left(\hat{a}^2+2 \hat{a}
   \sigma +\frac{27}{8}\right)-\frac{7}{8} x^2 (8 \hat{a}+3 \sigma ) \nonumber \\ &+x^{5/2} \left(\frac{26 \hat{a}^2}{9}+\frac{4 \hat{a}
   \sigma }{3}+\frac{135}{16}\right)+\frac{3}{16} x^3 \left(8 \hat{a}^2 \sigma -108 \hat{a}-27 \sigma
   \right)+O\left(x^{7/2}\right), \label{eq:JMPD} \\
p_\varphi^{\rm EOB} &= \frac{1}{\sqrt{x}}+\frac{3 \sqrt{x}}{2}-\frac{5}{6} x (4 \hat{a}+3 \sigma )+x^{3/2} \left(\hat{a}^2+2 \hat{a} \sigma
   +\frac{27}{8}\right)-\frac{7}{8} x^2 (8 \hat{a}+3 \sigma ) \nonumber \\ &+x^{5/2} \left(\frac{26 \hat{a}^2}{9}+\frac{4 \hat{a} \sigma
   }{3}+\frac{135}{16}\right)-\frac{9}{16} x^3 \left(8 \hat{a}^2 \sigma +36 \hat{a}+9 \sigma
   \right)+O\left(x^{7/2}\right). \label{eq:pphiEOB}
\end{align}
\end{widetext}

\begin{figure}[htp!]
\includegraphics[width=0.45\textwidth]{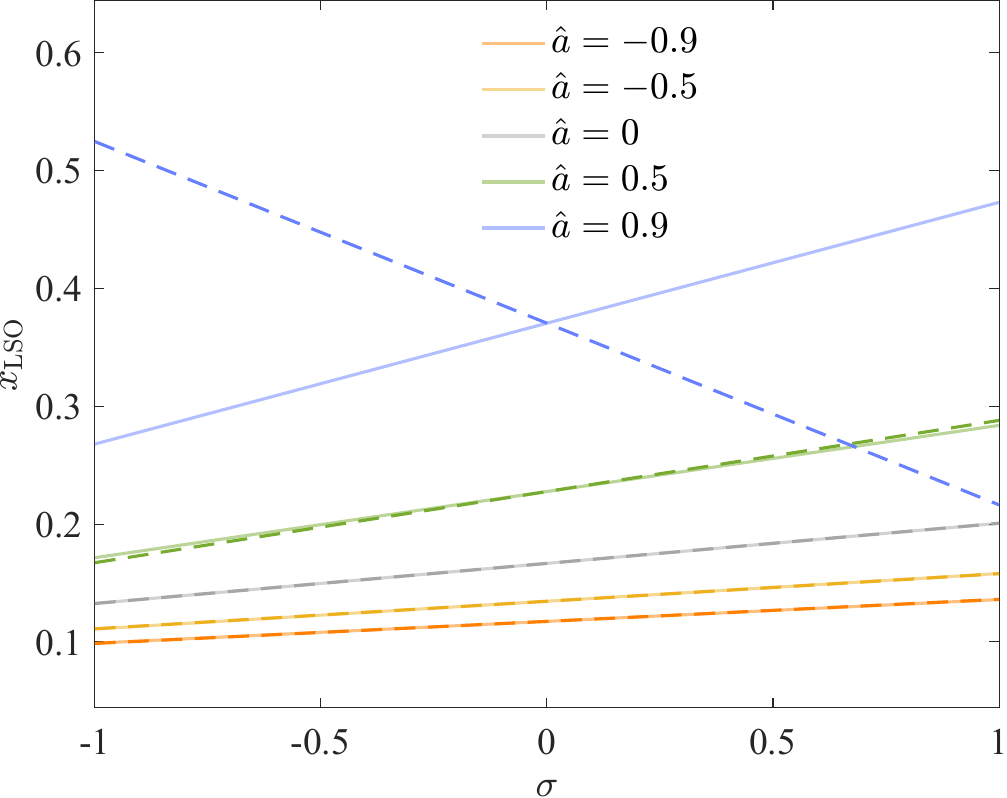} \\
\includegraphics[width=0.45\textwidth]{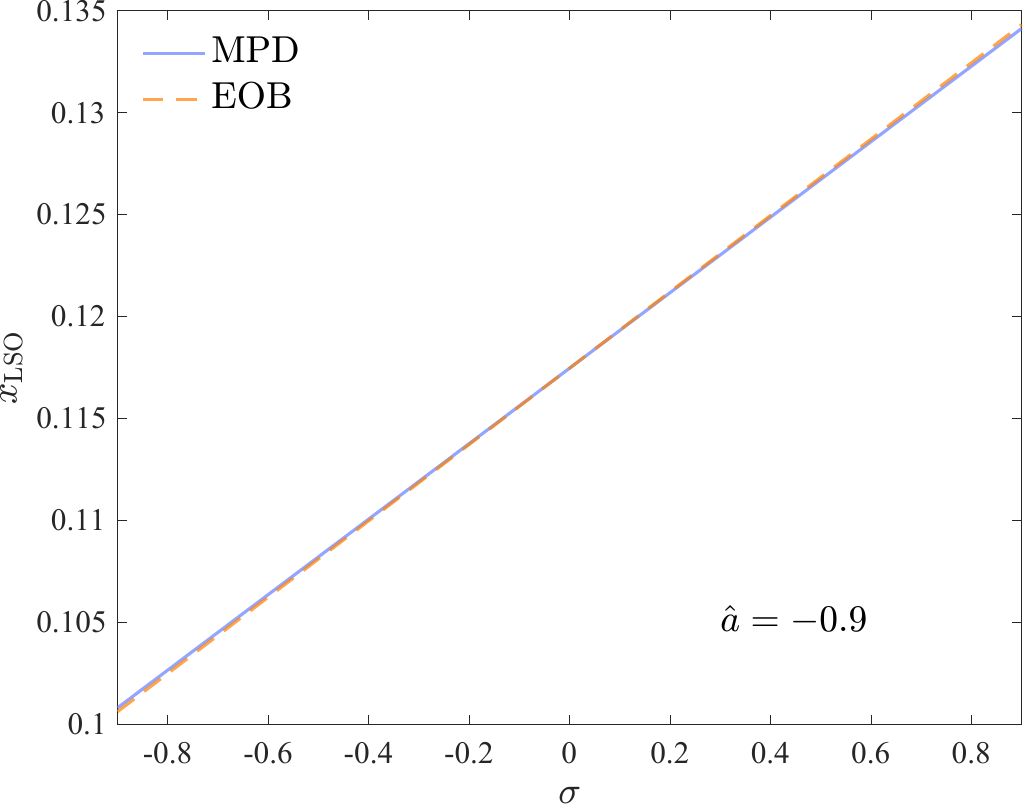}
\caption{\label{fig:LSO} \textit{Top}: Values of $x$ at the LSO for $\hat{a} = \{-0.9, -0.5, 0., 0.5, 0.9 \}$ as a function of $\sigma$, where the solid and dashed lines correspond respectively to MPD and EOB. The values are obtained linearizing in $\sigma$ as described in the text. The weird behavior of the EOB curve for $\hat{a} = 0.9$ is clarified when computing the non-linearized LSO (see Fig.~\ref{fig:EOBLSO}). \textit{Bottom}: Focus on $\hat{a} = -0.9$, showing that for this case the EOB $x_{\rm LSO}$ is larger/smaller than the MPD one for positive/negative $\sigma$. In the following we will see that this corresponds to the EOB asymptotic fluxes being smaller/larger than the MPD ones.}
\end{figure}

The difference between Eq.~\eqref{eq:JMPD} and Eq.~\eqref{eq:pphiEOB} is $\sigma +6 \hat{a}^2 \sigma  x^3+O\left(x^{7/2}\right)$. We thus add $\sigma$ to the EOB orbital angular momentum in comparing it to the MPD quantity obtained using Eq.~\eqref{eq:CEOJz}. Furthermore, we see that the rest of the difference between the two expansions is consistent with the fact that spin-orbit couplings in this EOB Hamiltonian are reproduced up to 2.5PN. We note here that in order to reach this agreement it was necessary to include the NNLO contribution into the EOB centrifugal radius.

Table~\ref{tab:consts} displays the energy and the angular momentum for EOB and MPD for values of the frequency parameter corresponding to $x = \{ 0.05, 0.1, 1/6, 0.2\}$, for $\hat{a} = \{ -0.9, 0.9\}$ and $\sigma = \{ -0.9, 0., 0.9 \}$. Grey values are beyond the LSO, whose computation will be presented in the next section. We note that for nonspinning secondary, as expected, the EOB and MPD values coincide. On the other hand, for positive values of $\sigma$, the EOB values are always smaller than the MPD ones, while for negative values of $\sigma$ the inverse is true. In both cases, the difference is larger when $\hat{a}$ is positive.

\begin{figure*}[htp!]
\includegraphics[width=0.42\textwidth]{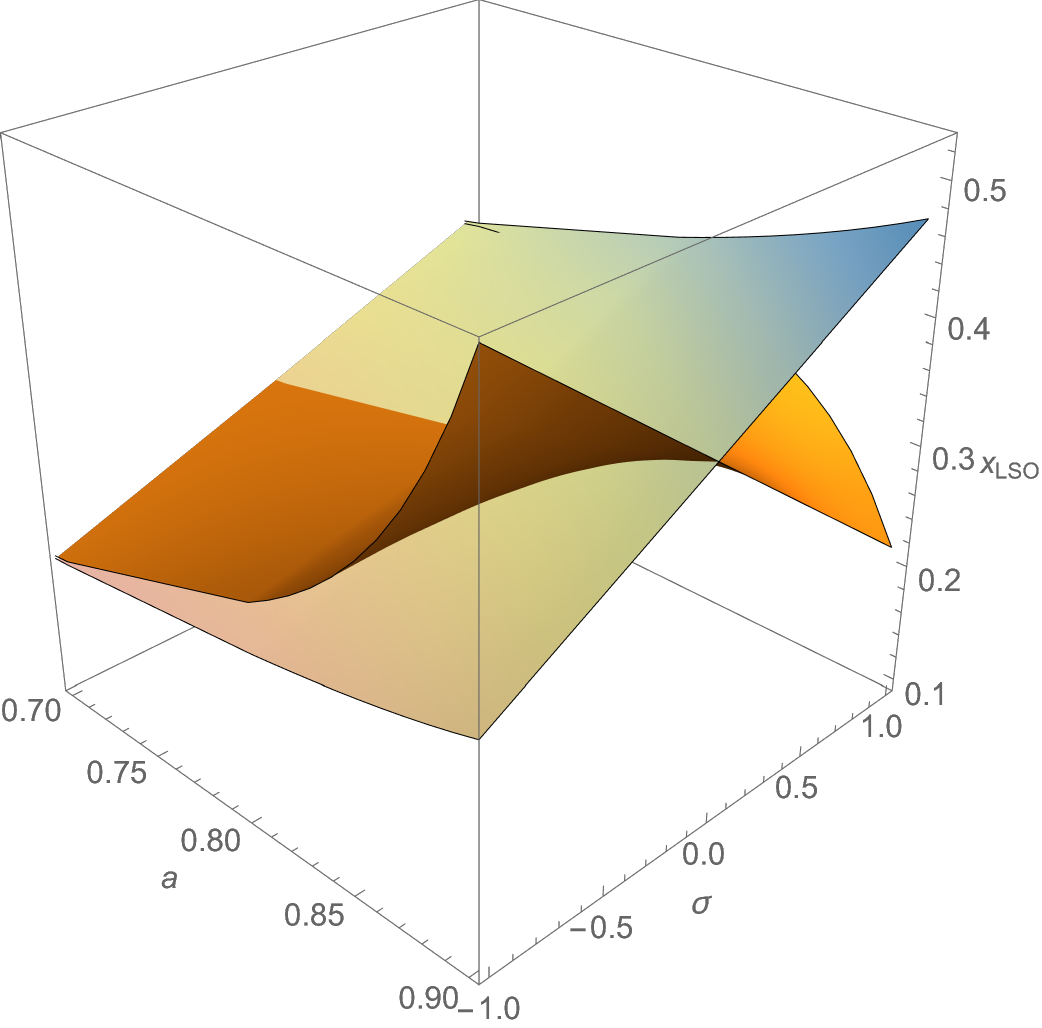} 
\hspace{0.5mm}
\includegraphics[width=0.42\textwidth]{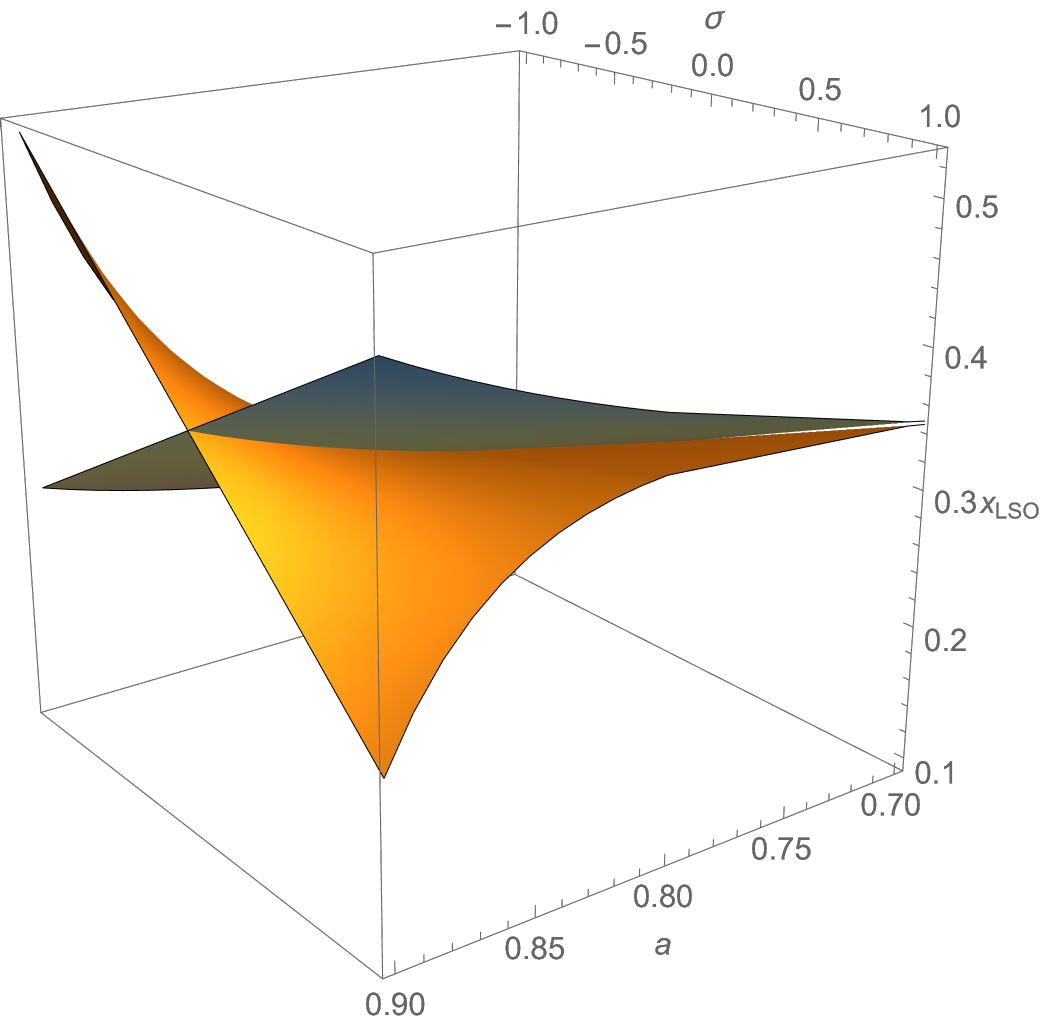} 
\caption{\label{fig:3D} Behavior of $x$ at the LSO as a function of $\{ \hat{a}, \sigma\}$, in the range $\hat{a} = [0.7,0.9]$. For both EOB (orange) and MPD (pastel) $x$ is linearized in $\sigma$ as described in the text. As $\hat{a}$ increases, the EOB $x_{\rm LSO}$ inverts its proportionality to $\sigma$, explaining what we see in Fig.~\ref{fig:LSO}.}
\end{figure*}

\begin{figure}
\includegraphics[width=0.45\textwidth]{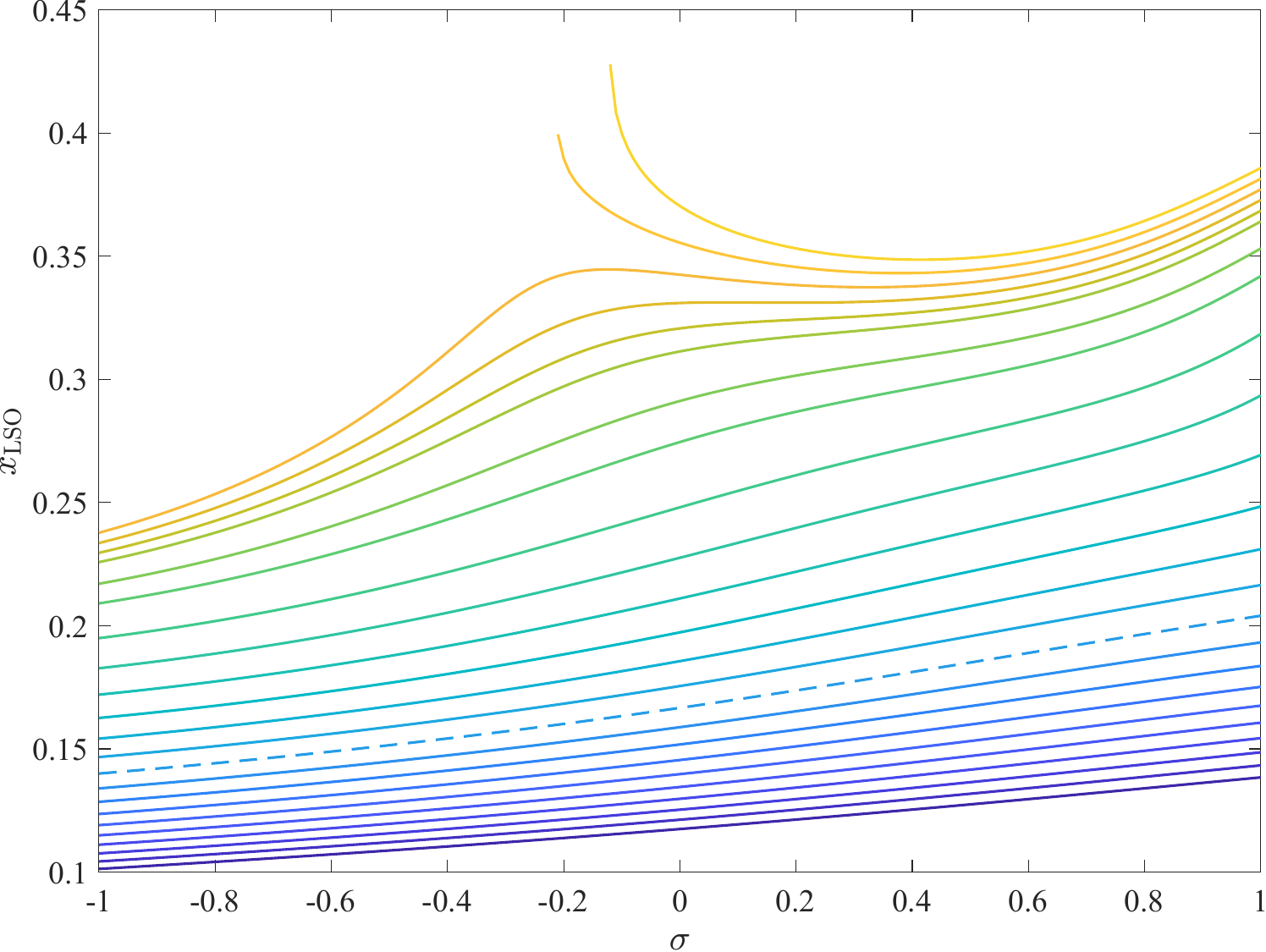} 
\caption{\label{fig:EOBLSO} Values of $x$ at the LSO as a function of $\sigma$ for $\hat{a}$ in the range $[-0.9,0.9]$, evaluated from the EOB dynamics without linearizing in spin. The lines are related to growing values of $\hat{a}$ from bottom to top, and the dashed one is for $\hat{a} = 0$. The plot shows how the full EOB dynamics does not have a LSO below $\sigma \simeq -0.2$ for large, positive values of $\hat{a}$.}
\end{figure}

The fact that the EOB energy is larger/smaller than the MPD one at a given frequency for $\sigma < 0 / \sigma > 0$ suggests that the EOB spin-orbit interaction is stronger than the MPD one for these values of $\{ \hat{a}, \sigma \}$. Furthermore, for these values of $\hat{a}$, negative secondary spins yield a larger energy than positive ones, which is again due to spin-orbit interaction. We have verified however that for different values of $\{ \hat{a}, \sigma \}$ the aforementioned behaviors change, so that the results cannot be simply ascribed to spin-orbit interaction, and spin-spin contributions play a more important role. 

\section{LSO}
\label{sec:LSO}

Before presenting the numerical fluxes, in this section we evaluate LSO $x$ values for both MPD and EOB.

For the MPD case, we consider again the expression for the orbital frequency $\Omega$ shown in Eq.~(22) of Ref.~\cite{Harms:2015ixa}, and expand $x = \Omega^{2/3}$ at linear order in $\sigma$, which yields
\begin{align}
x(u) &= x_0(u) + \sigma x_1(u) =\\
               &= \left( \frac{u^{3/2}}{1 + a u^{3/2}} \right)^{2/3}  - \sigma \left( \frac{u^{3/2}}{1 + a u^{3/2}} \right)^{5/3} (1 - a u^{1/2})
\end{align}
where $u$ here corresponds to $u_{\rm MPD}$. We then exploit the LSO expression $u_0 + \sigma u_1$ from Ref.~\cite{Jefremov:2015gza}, where $u_0$ is found as the inverse of the root of the equation
\be
r^2 - 6r - 3 a^2 + 8 a \sqrt{r} = 0
\ee
and 
\be
u_1 = -4 u_0^2 (a u_0 - \sqrt{u_0}) .
\ee
We substitute $u_0 + \sigma u_1$ into $x_0$ and further linearize the term in $\sigma$, and in $x_1$ we only substitute $u_0$. 

For  EOB, the LSO values for $u$ and $p_\varphi$ are found by imposing $\p_u H_{\rm eff} = \p_u^2 H_{\rm eff} = 0$. In this case, we substitute $u = u_0 + \sigma u_1$ and $p_\varphi = p_{\varphi,0} + \sigma p_{\varphi,1}$ into $\p_u H_{\rm eff}$ and $\p_u^2 H_{\rm eff}$ and further linearize these expressions in $\sigma$. Then, for given $\hat{a}$ we exploit the values $\{u_0, p_{\varphi,0}\}$ for a nonspinning particle, and solve for $u_1$ and $p_{\varphi,1}$. We do this within a loop on $\sigma$ values with $\Delta \sigma = 0.01$, and use the Mathematica routine \texttt{FindRoot}, exploiting as initial guess for every value of $\sigma$ the solution of the previous step.

\begin{figure}
\includegraphics[width=0.48\textwidth]{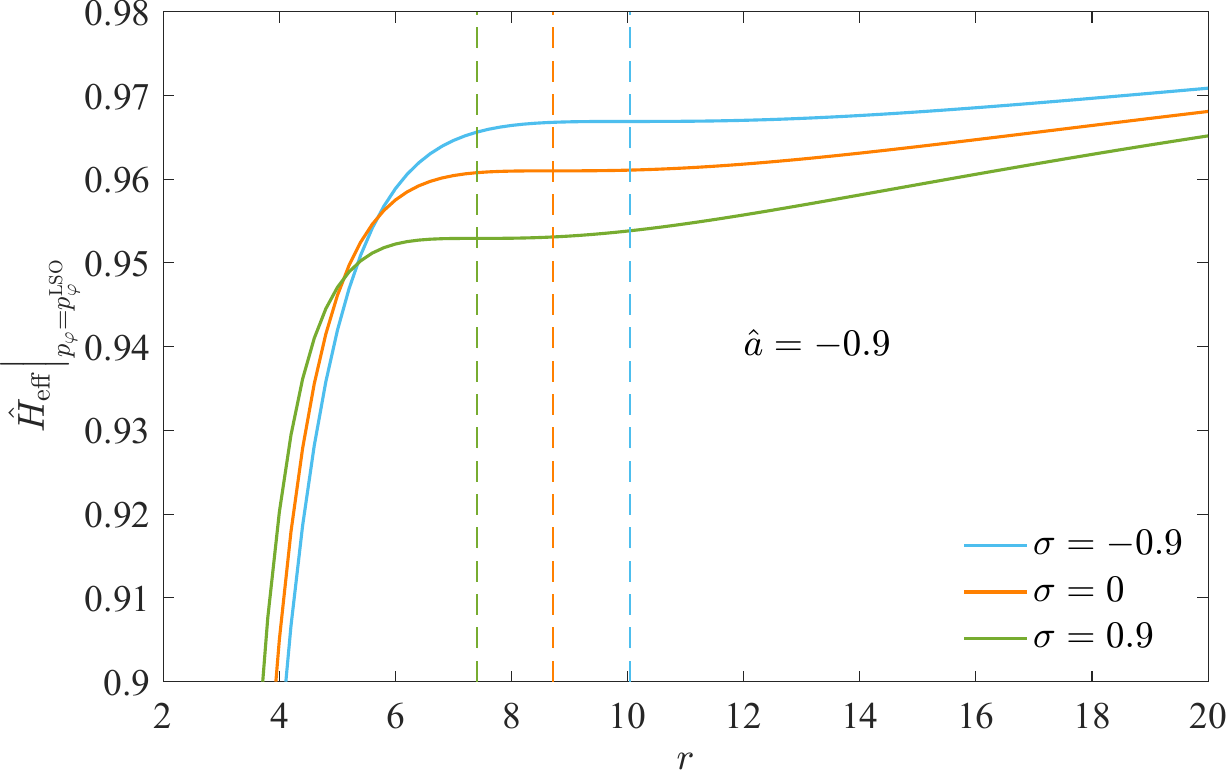} 
\caption{\label{fig:HeffAtLSO} The EOB Hamiltonian evaluated at $p_\varphi^{\rm LSO}$ for $\hat{a} = -0.9$ and $\sigma = \{ -0.9, 0, 0.9\}$. Vertical dashed lines indicate values of the LSO radius. We can see how spin-orbit interaction due to the spin of the secondary causes together an energy shift and a LSO shift.}
\end{figure}

The $x_{\rm LSO}$ values for EOB and MPD are presented in the upper plot in Fig.~\ref{fig:LSO}, for $\hat{a} = \{ -0.9, -0.5, 0., 0.5, 0.9\}$. Fig.~\ref{fig:LSO} also displays in the bottom plot a focus on the case for $\hat{a} = -0.9$, for which the EOB $x_{\rm LSO}$ is larger/smaller than the MPD one for positive/negative $\sigma$. As discussed above for the energy values, we again ascribe the LSO differences to spin-orbit interaction. In fact, spin-orbit interaction yields a positive/negative energy shift when the spin is anti-aligned/aligned with the orbital angular momentum. In Fig.~\ref{fig:HeffAtLSO} we plot the (non-linearized-in-$\sigma$) EOB Hamiltonian evaluated at $p_\varphi^{\rm LSO}$, showing how a larger energy corresponds to a larger LSO radius. While spin-spin contributions are also present, at least for this value of $\hat{a}$ we can interpret this evidence simply by spin-orbit interaction. In the following we will see that this also corresponds to the EOB asymptotic fluxes being smaller/larger than the MPD ones for positive/negative $\sigma$.

We also find an unexpected behavior of the LSO for $\hat{a} = 0.9$. To better understand this, we compute $x_{\rm LSO}$ for more values of $\hat{a}$ and plot the EOB and MPD behaviors as function of $\hat{a}$ and $\sigma$ in Fig.~\ref{fig:3D}. The plots show how the EOB $x_{\rm LSO}$ is changing its dependence on $\sigma$ as $\hat{a}$ becomes large and positive. For completeness, in the supplementary material~\cite{supp} we attach the complete 3D picture in the range $\hat{a} = \{-0.9, 0.9\}$.

To gain a better understanding of this behavior, we also compute the EOB non-linearized values, obtained simply solving $\p_u H_{\rm eff} = \p_u^2 H_{\rm eff} = 0$ for $\{u, p_\varphi \}$ and considering $x = \Omega^{2/3} = (\p_{p_\varphi} H_{\rm eff})^{2/3}$ without linearizing. Our result is shown in Fig.~\ref{fig:EOBLSO}, and explains that what we witness in the linearized case is due to an absence of the EOB LSO for large, positive $\hat{a}$ and $\sigma \lesssim -0.2$.

It is a well-known fact that the comparable-mass version of \TEOBResumS{} lacks an LSO for large, positive spins. Ref.~\cite{Rettegno:2019tzh} showed that incorporating the spinning-particle $G_{S_*}$, using the EOB potentials instead of the Kerr functions $A_{\rm eq}, B_{\rm eq}$, and thus working in the anti-DJS gauge, allowed the model to have a LSO for every value of the spin. Our computation shows however that using the complete spinning-particle $G_{S_*}$ in the test-mass limit does not solve the issue. It will be interesting in the future to investigate whether there are options for the EOB spin gauge that allow the model to have an LSO for every value of the spins, both in the comparable-mass and in the test-mass regimes. We leave this analysis to future work.

\section{Fluxes}
\label{sec:fluxes}

In this section we present the fluxes evaluated with the FD TE solver of Ref.~\cite{Skoupy:2021asz}. The solver takes as input (i) the values of $\hat{a}$ and $\sigma$, (ii) the BL radius, (iii) the azimuthal frequency $\Omega$ (iii) the energy and the angular momentum of the orbit. We note here that this code is \textit{not} linearized in the secondary spin. We evaluate the fluxes for fixed values of $x = [ 0.001, 0.201]$ with steps $\Delta x = 0.02$. For these values, the $u(x)$ relations we derived above are used to compute the energy and angular momentum for EOB and MPD at a given $x$. We consider here just the energy fluxes, since for circular orbits the relation $\dot{E} = \dot{L}_z/\Omega$ holds. We sum over the modes $\ell = 2, 3$ and $m = 1, 2, 3$. When plotting the fluxes, we normalize them by the leading-order Newtonian contribution, which for the infinity flux reads
\be
\dot{E}_{\rm Newt}^{\infty} = \frac{32}{5} x^5 .
\ee
As for the horizon flux, for nonspinning binaries it holds
\be
\label{eq:NewtHFnonspin}
\dot{E}_{\rm Newt, ns}^{\mathcal{H}} = \dot{E}_{\rm Newt}^{\infty} x^4 ,
\ee
while for spinning binaries we consider the LO contribution in the expression\footnote{Note that the authors of Ref.~\cite{Saketh:2022xjb} define $x \equiv (M\Omega)^{1/3}$.} in Eq.~(4.26) of Ref.~\cite{Saketh:2022xjb}, which reads
\be 
\label{eq:NewtHFspin}
\dot{E}_{\rm Newt, s}^{\mathcal{H}} = \dot{E}_{\rm Newt, ns}^{\mathcal{H}} \left( -\frac{a(1 + 3 a^2)}{4 \Omega}  \right) ,
\ee

Note that when considering the linear-in-$\sigma$ contribution, namely
\be
\dot{E}^\mathcal{H} = \dot{E}^\mathcal{H}_0 + \sigma \dot{E}^\mathcal{H}_1 ,
\ee
where $\dot{E}^\mathcal{H}_0$ is
\be
\dot{E}^\mathcal{H}_0 = \dot{E}_{\rm Newt, ns}^{\mathcal{H}} x^{-3/2} \left( -\frac{a(1 + 3 a^2)}{4} + \mathcal{O}(x) \right) ,
\ee
we have found numerically that
\be\label{eq:Edot^H_1}
\dot{E}^\mathcal{H}_1 = \dot{E}_{\rm Newt, ns}^{\mathcal{H}} x^{-3/2} \left( -\frac{a(1 + 3 a^2)}{2} x^{3/2} + \mathcal{O}(x^2) \right) .
\ee
In particular, we found this expression by numerically calculating the $l=m=2$ flux for very small $x$, taking the numerical derivative with respect to $\sigma$ and then fitting the dependence on the primary spin. Note that the linear-in-$\sigma$ part starts $3/2$ PN orders higher than the nonspinning-secondary part.

\begin{figure*}[htp!]
\includegraphics[width=0.45\textwidth]{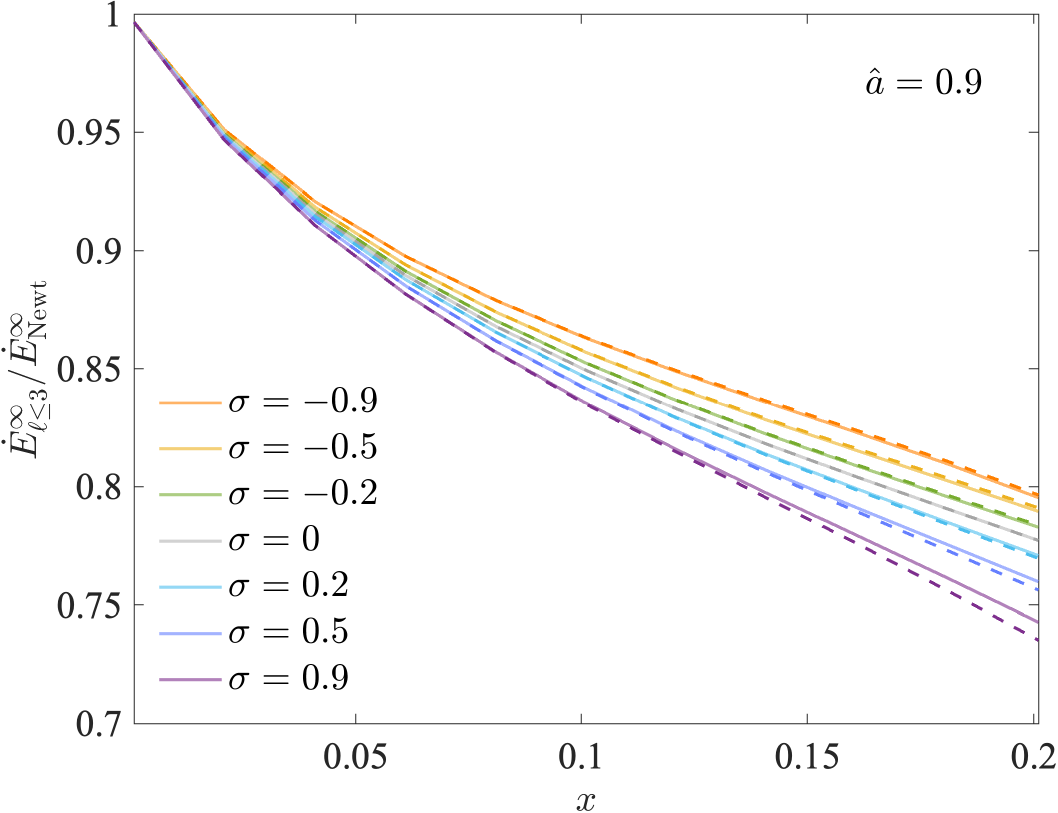} 
\includegraphics[width=0.45\textwidth]{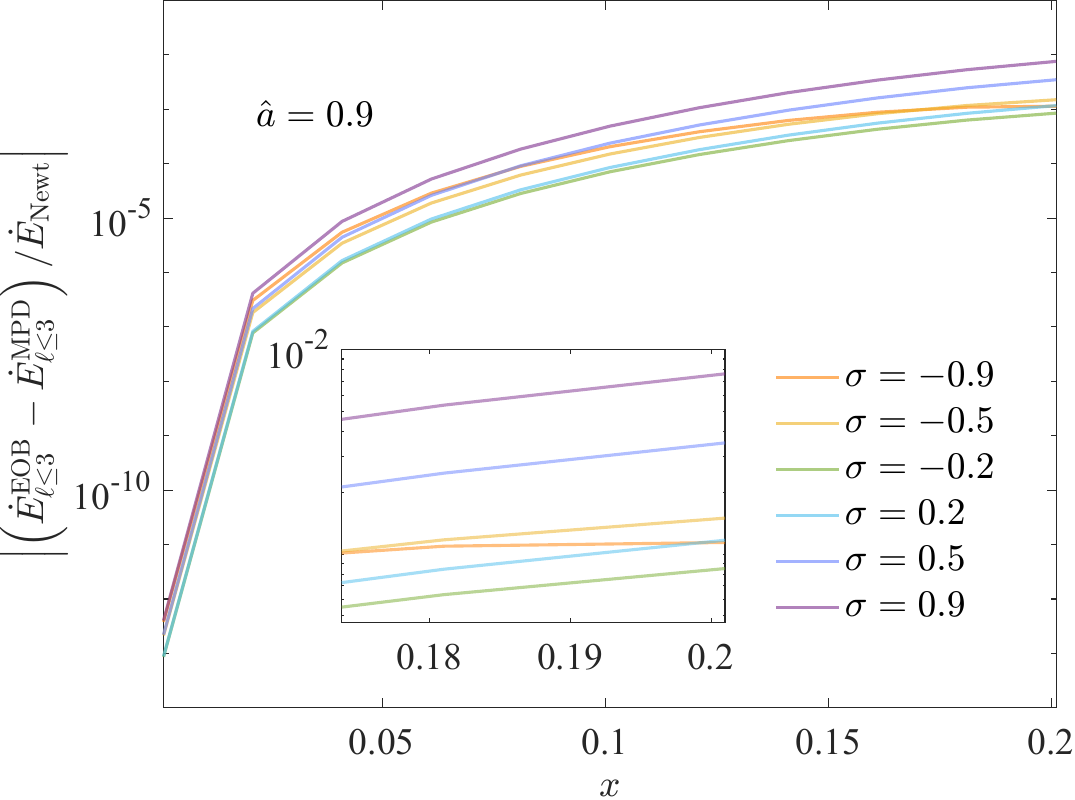} \\
\vspace{0.5mm}
\includegraphics[width=0.45\textwidth]{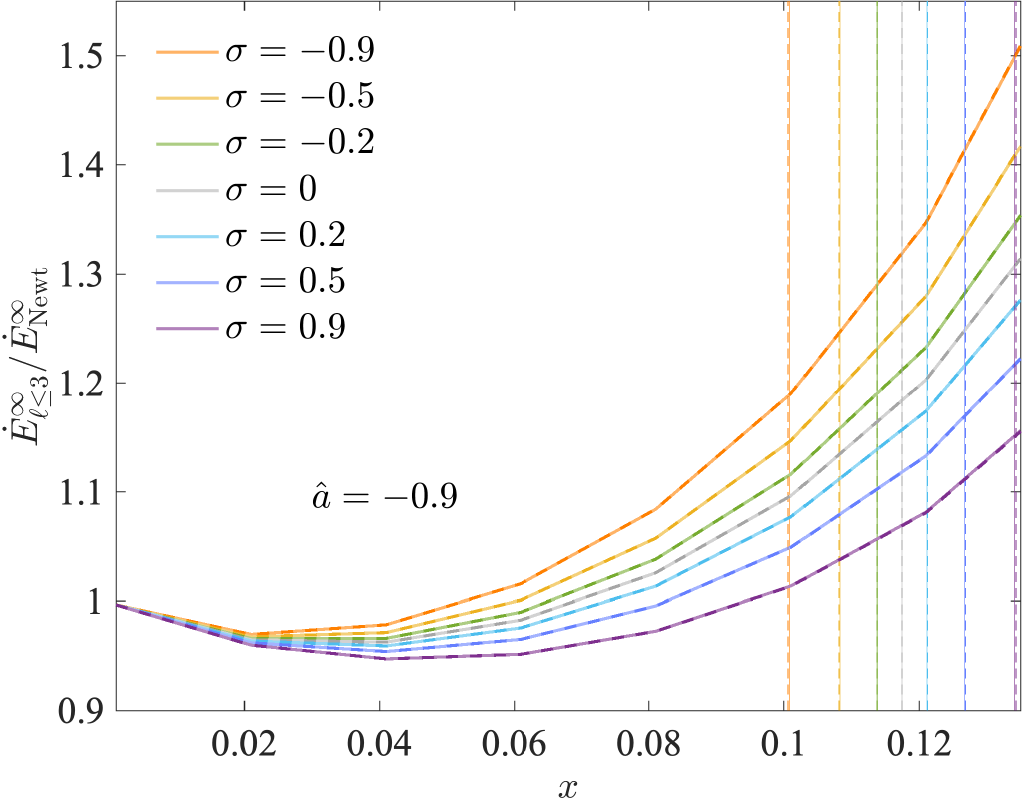}
\includegraphics[width=0.45\textwidth]{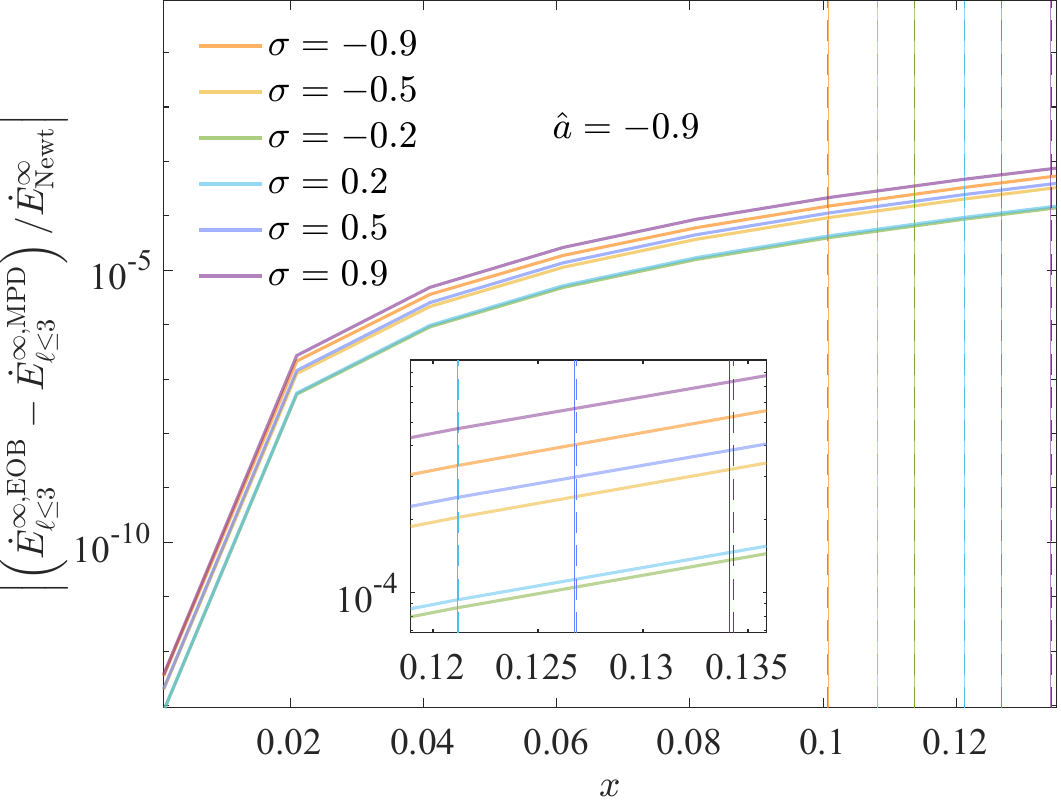}
\caption{\label{fig:fluxes_I} \textit{Left column:} Gravitational wave energy flux at infinity evaluated with the FD TE solver with the EOB and MPD dynamic values, for values of the Kerr parameter corresponding to $a = \{-0.9, 0.9\}$ respectively shown in the bottom and top rows. The flux is summed over $\ell = 2, 3$ and $m = 1, 2, 3$ and is normalized by the Newtonian contribution. In every plot, the solid lines correspond to the MPD values and the dashed lines to the EOB ones. For $\hat{a} = -0.9$ we also plot vertical lines for the LSO $x$ values, again solid/dashed for MPD/EOB. \textit{Right column:} Difference between the EOB and MPD fluxes normalized by the Newtonian contribution, again for the same aforementioned values of the Kerr parameter. We do not plot the curve for $\sigma = 0$ since it evaluates exactly (up to the considered digits) to zero.}
\end{figure*}

Fig.~\ref{fig:fluxes_I} shows the fluxes at infinity for MPD (solid lines) and EOB (dashed lines) for $\hat{a} = \{-0.9, 0.9\}$ and $\sigma = \{ -0.9, -0.5, -0.2, 0, 0.2, 0.5, 0.9\}$, together with the Newtonian-normalized difference between the EOB and MPD values. As for $\hat{a} = -0.9$, we actually consider the fluxes only up to the largest LSO value, which is $\sim 0.135$, and add vertical lines for the MPD and EOB LSO values (respectively solid and dashed). As for  $\hat{a} = -0.9$, $x_{\rm LSO}$ is larger than $0.201$ for all values of $\sigma$. The EOB asymptotic fluxes are larger than the MPD ones for negative spins, while the opposite holds for positive spins. As mentioned in the previous section, we find that, at least for $\hat{a} = -0.9$, this is related to the EOB LSO radius being larger/smaller than the MPD one for negative/positive values of $\sigma$, and correspondingly to the EOB energy values being larger/smaller than the MPD ones. We attribute this to the EOB spin-orbit interaction being stronger than the MPD one.

\begin{figure*}[htp!]
\includegraphics[width=0.45\textwidth]{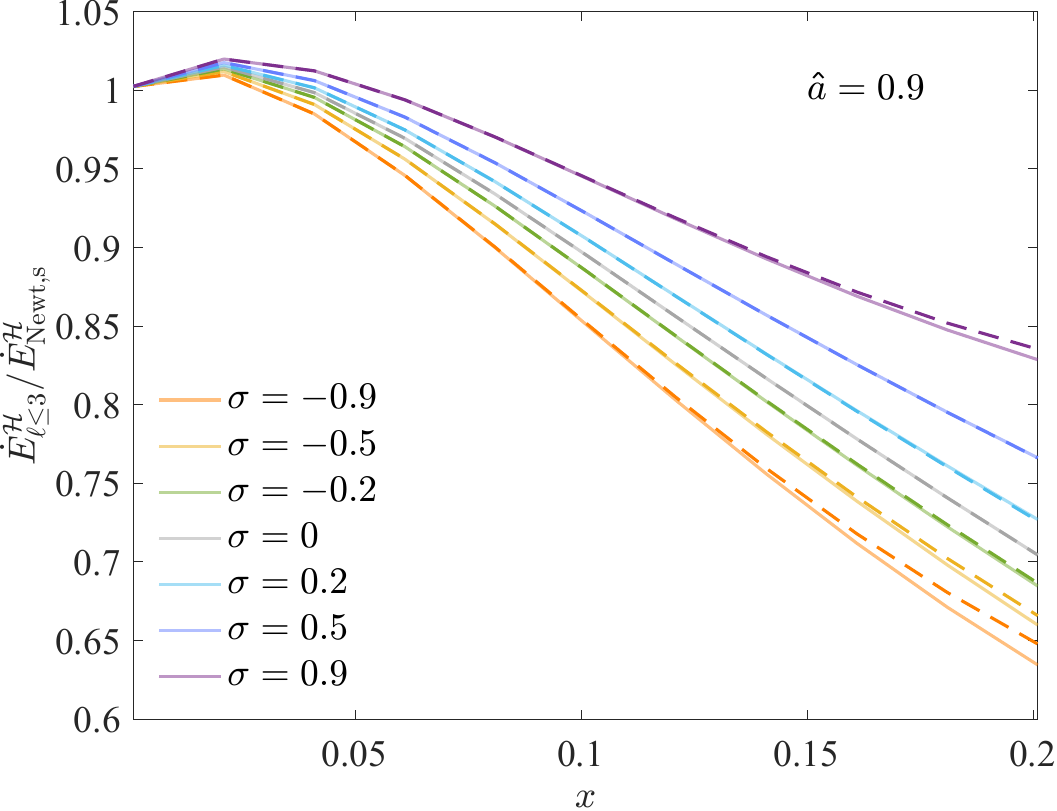} 
\includegraphics[width=0.45\textwidth]{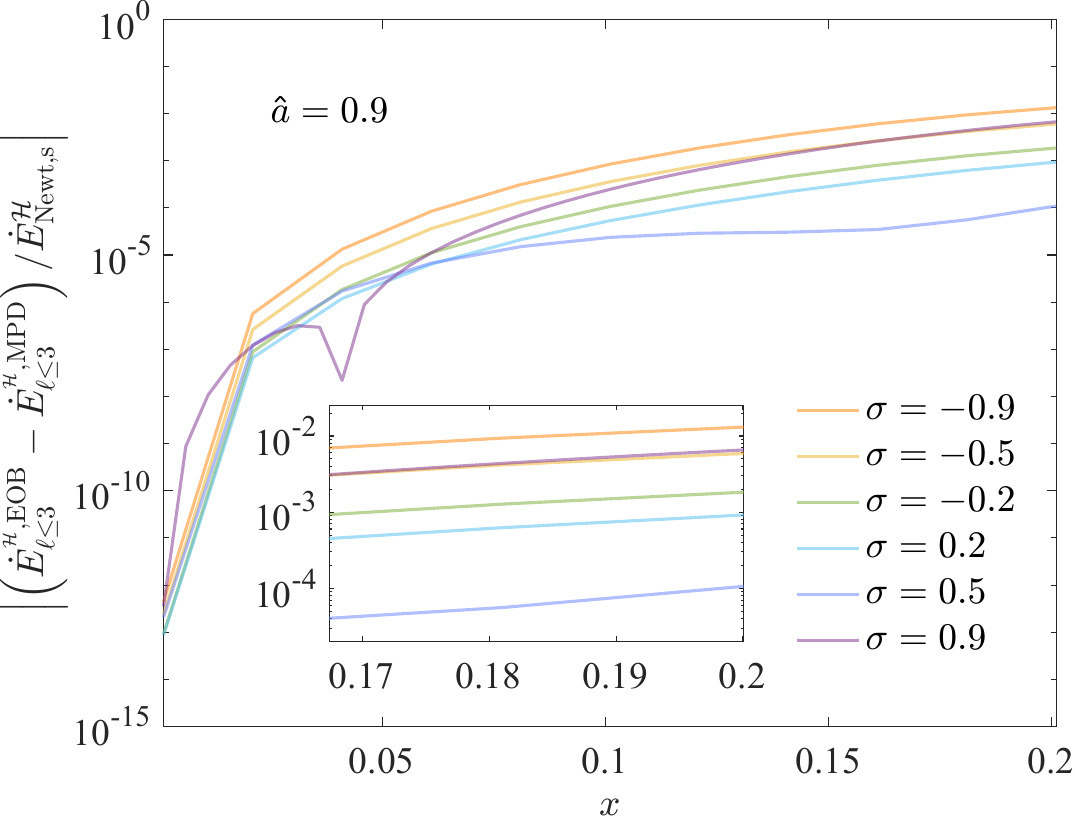} \\
\vspace{0.5mm}
\includegraphics[width=0.45\textwidth]{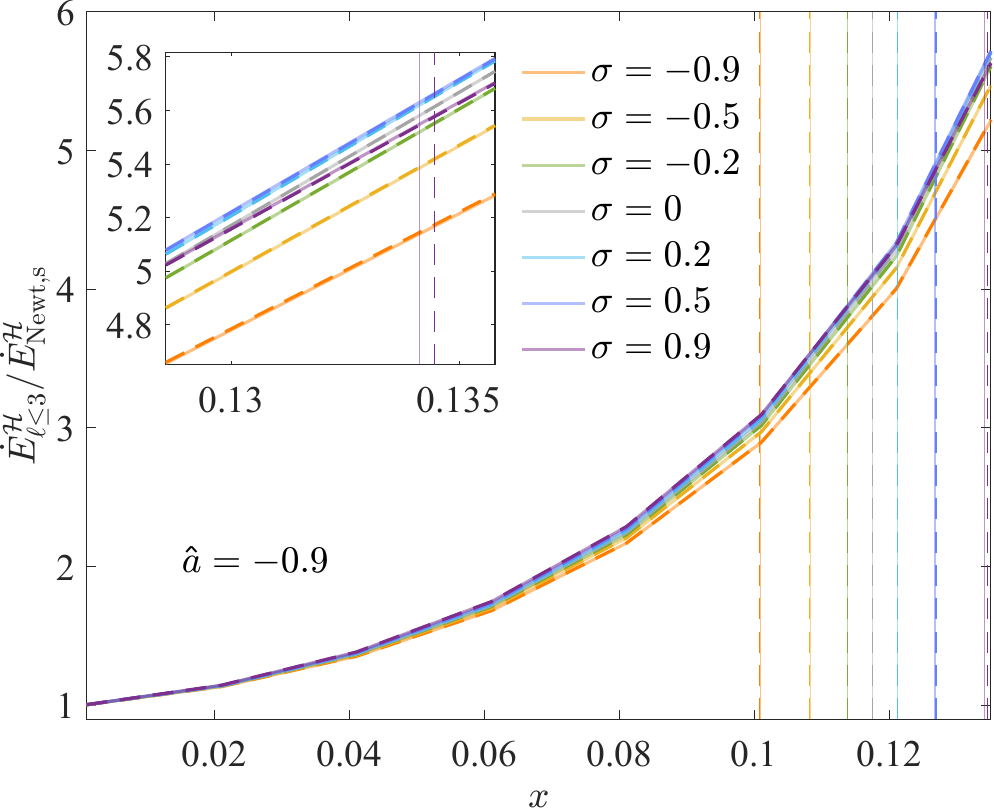}
\includegraphics[width=0.45\textwidth]{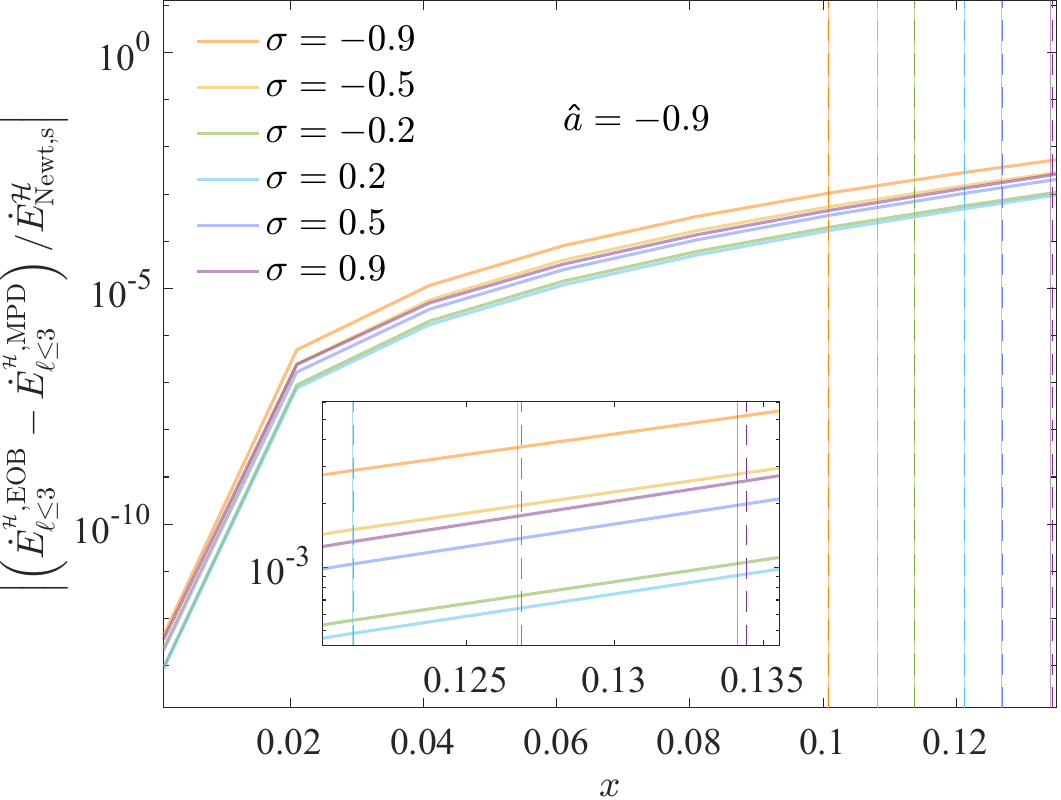}
\caption{\label{fig:fluxes_H} \textit{Left column:} Gravitational wave energy flux at the horizon evaluated with the FD TE solver with the EOB and MPD dynamic values, for values of the Kerr parameter corresponding to $\hat{a} = \{-0.9, 0.9\}$ respectively shown in the bottom and top rows. The flux is summed over $\ell = 2, 3$ and $m = 1, 2, 3$ and is normalized by the Newtonian contribution. In every plot, the solid lines correspond to the MPD values and the dashed lines to the EOB ones. For $\hat{a} = -0.9$ we also plot vertical lines for the LSO $x$ values, again solid/dashed for MPD/EOB. \textit{Right column:} Difference between the EOB and MPD fluxes normalized by the Newtonian contribution, again for the same aforementioned values of the Kerr parameter. We do not plot the curve for $\sigma = 0$ since it evaluates exactly (up to the considered digits) to zero.}
\end{figure*}

Fig.~\ref{fig:fluxes_H} is the analogous of Fig.~\ref{fig:fluxes_I} for the horizon fluxes. The EOB fluxes are still larger/smaller than the MPD ones for negative/positive spins, apart from the case $\{\hat{a}, \sigma\} = \{0.9, 0.9\}$. For this case, when plotting the EOB/MPD difference on a logarithmic scale in the upper right panel of Fig.~\ref{fig:fluxes_H}, we compute denser point in $x$, with $\Delta x = 0.005$. This allows us to see the singularity in the logarithmic difference, which signals a change in the absolute difference, namely indicates that the EOB horizon flux for $\{\hat{a}, \sigma\} = \{0.9, 0.9\}$ is smaller than the MPD one before $x \sim 0.05$ and larger than the MPD one after that. We are not sure of the exact reason behind this behavior, but we suspect that the EOB linearized-in-$\sigma$ EOB description might have unphysical features for these values of $\{ \hat{a}, \sigma\}$.

The horizon flux for $\{\hat{a}, \sigma\} = \{-0.9, 0.9\}$ for both EOB and MPD also has a peculiar behavior, which is highlighted in the small panel in the bottom left plot Fig.~\ref{fig:fluxes_H}. This might be due to the fact that the quadratic-in-spin contribution to the fluxes is large for such large values of $\sigma$, but the TE solver is fed with a linearized-in-$\sigma$ dynamics, which makes the quadratic contribution to the fluxes incomplete. However, in Appendix~\ref{sec:PN_HF} we see that a similar behavior is found when considering an analytical PN formula for the horizon flux. Intuitively, we can ascribe this to an interplay between spin-orbit and spin-spin effects.

It is also curious to notice that as for the horizon flux, positive/negative values of $\sigma$ are related to a larger/smaller emission than the $\sigma = 0$ case, which is the opposite of what holds for the asymptotic fluxes. We have confirmed this finding by using a time domain Teukolsky equation solver called {\rm Teukode} \cite{Harms:2014dqa} to calculate the horizon fluxes. A more detailed discussion regarding fluxes through the horizon is provided in Appendix~\ref{sec:PN_HF}.

In Fig.~\ref{fig:fluxes_a0} we plot both the infinity and the horizon fluxes for $a = 0$. These show that for a nonspinning primary, at linear order in the secondary spin, EOB and MPD give the same result, namely the difference between the EOB and MPD values evaluates exactly (up to the considered digits) to zero for all the considered values of $x$ and $\sigma$.

We note here that Ref.~\cite{Harms:2016ctx} had found a difference in the asymptotic fluxes for a spinning test particle on a Schwarzschild BH that were evaluated using the EOB dynamics and the MPD one with different SSCs. This was due to the fact that the dynamics of Ref.~\cite{Harms:2016ctx} was not linearized in $\sigma$. In Appendix~\ref{sec:Schwarzschild} we give some details about this computation, and show that the EOB and MPD (under TD SSC) values for the energy for a spinning particle on CEOs around a Schwarzschild BH agree if evaluated analytically at linear order in $\sigma$. This implies that when linearizing in $\sigma$ for a nonspinning primary, the dynamics fed to the TE solver is the same for EOB and MPD under TD SSC and thus there is no difference in the evaluated fluxes.

\begin{figure*}
\includegraphics[width=0.45\textwidth]{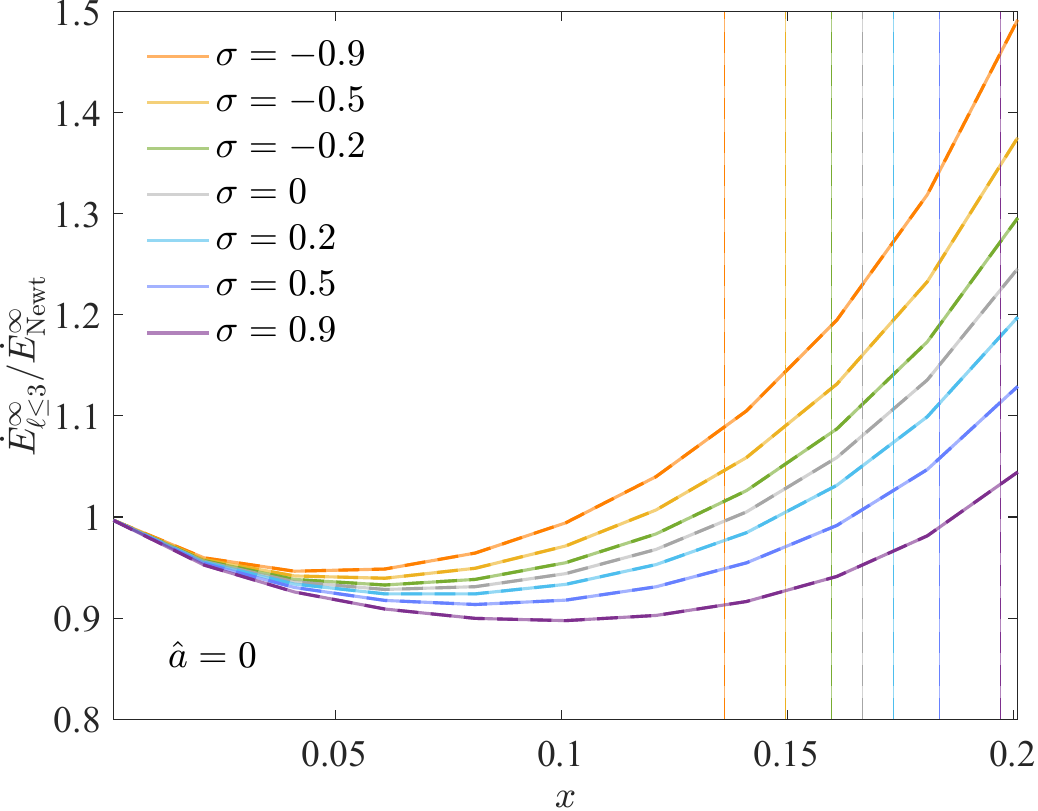} 
\includegraphics[width=0.45\textwidth]{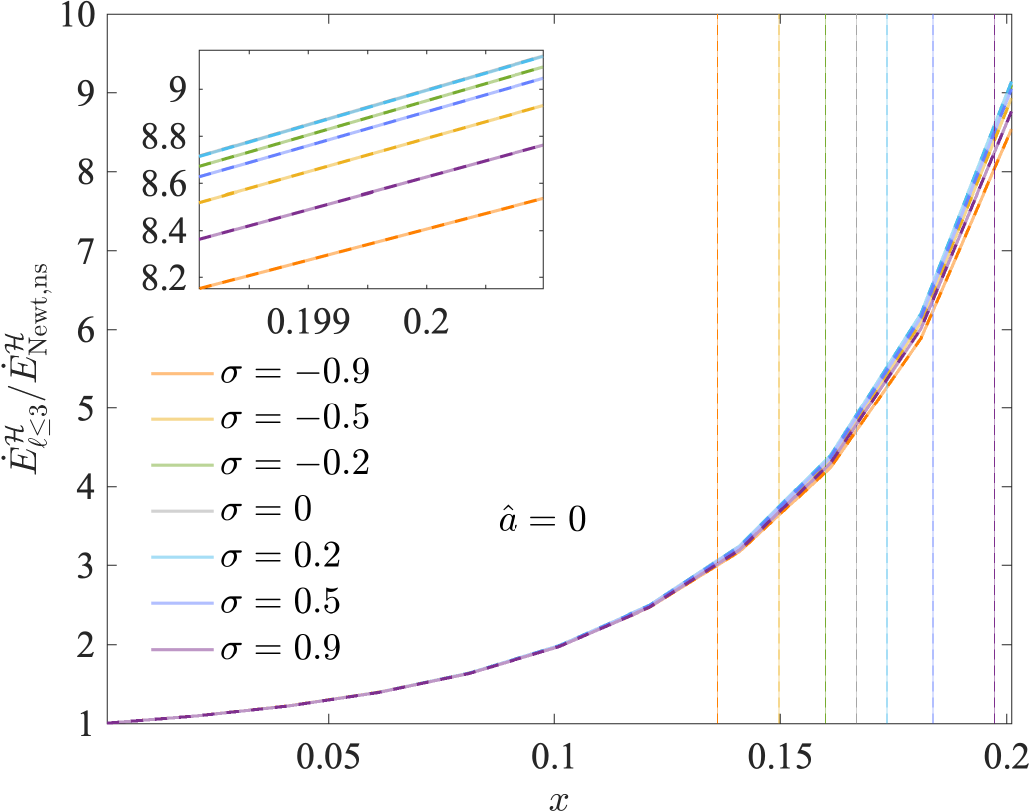} 
\caption{\label{fig:fluxes_a0} Gravitational wave energy fluxes at infinity (left) and at the horizon (right) for evaluated with the FD TE solver with the EOB and MPD dynamic values, for a nonspinning primary. The flux is summed over $\ell = 2, 3$ and $m = 1, 2, 3$ and is normalized by the Newtonian contribution. In every plot, the solid lines correspond to the MPD values and the dashed lines to the EOB ones. We also plot vertical lines for the LSO $x$ values, again solid/dashed for MPD/EOB. Note that in the plot on the right, the curves for $\sigma  = 0$ are almost superposed with the ones for $\sigma = 0.2$. We do not plot the difference between the EOB and MPD values in this case since it evaluates exactly (up to the considered digits) to zero for all the values of $x$ and $\sigma$ that we take into account.}
\end{figure*}

\section{Conclusions}
\label{sec:conclusions}

In this paper we have computed dynamical quantities for the EOB and MPD dynamics for a spinning particle on CEOs around a Kerr BH linearizing in the secondary spin. In particular, we have evaluated the relation between radius and frequency, and exploited it to have linearized-in-$\sigma$ expressions for the energy and angular momentum in terms of the frequency parameter $x$. The latter have been fed to a Teukolsky equation frequency-domain solver to evaluate the asymptotic and horizon gravitational-wave fluxes for values of the Kerr parameter $\hat{a} = \{-0.9, 0.9\}$ and secondary spin $\sigma = \{ -0.9, -0.5, -0.2, 0., 0.2, 0.5, 0.9\}$. We summarize here our main findings and remarks:
\begin{itemize}
    \item When comparing analytical expressions for the angular momentum as a function of $x$, we find a difference at 3PN between the EOB and the MPD function. This is consistent with the fact that spin-orbit couplings in the EOB Hamiltonian that we use in this work are reproduced up to 2.5PN.
    
    \item We find that for $\hat{a} = \{-0.9, 0.9\}$, the EOB energy is larger than the MPD one when $\sigma < 0$, which is related to a larger LSO radius (viceversa for $\sigma > 0$). We ascribe this to the EOB spin-orbit interaction being stronger than the MPD one for these values of $\hat{a}$.
    
    \item We compute the linearized-in-$\sigma$ values for $x_{\rm LSO}$ for both EOB and MPD and verify that for $\hat{a} = -0.9$, the EOB $x_{\rm LSO}$ is smaller/larger than the MPD one for negative/positive sigma, which corresponds to a larger/smaller LSO radius. We also find that the EOB values for the linearized-in-$\sigma$ $x_{\rm LSO}$ have the opposite behavior than expected for $\hat{a} = 0.9$. By also computing the non-linearized values, we find that this is actually due to the fact the EOB dynamics does not have a LSO for large, positive $\hat{a}$ and $\sigma \lesssim -0.2$. We leave to future work an investigation on which choice for the EOB spin-orbit sector could allow for the LSO existence for every value of the spins in both the comparable-mass and test-mass regimes.

    \item We presented the asymptotic and horizon fluxes evaluated via the FD TE solver fed with the EOB and MPD dynamics. Both for the asymptotic and the horizon fluxes, we find that the EOB values are larger/smaller than the MPD ones for negative/positive $\sigma$, which we ascribe again to the EOB spin-orbit interaction being stronger than the MPD one. We also find that for asymptotic fluxes, negative/positive values of $\sigma$ are related to a larger/smaller emission with respect to the nonspinning particle case, while the opposite is true for the horizon fluxes. While the result for the asymptotic fluxes is more intuitively understandble in terms of spin-orbit interaction, the result for the horizon fluxes suggests that spin-spin contributions play a larger role in this case. We expand on this in Appendix~\ref{sec:PN_HF}.

    \item We find that the EOB and MPD dynamics (and fluxes) coincide if the primary is a Schwarzschild BH. We explain how and why our result differs from what found previously in Ref.~\cite{Harms:2016ctx}. In particular, our results are due to a linearization in $\sigma$ for every quantity that we compute. This is explained in detail in Appendix~\ref{sec:Schwarzschild}.

\end{itemize}

Our work represents an extension to what had been presented in Ref.~\cite{Harms:2016ctx} for a spinning particle on Schwarzschild to a Kerr background. In our case, the use of a FD TE solver allows us to find more precise results for the fluxes. The computations presented here will be useful in benchmarking analytical choices for the EOB radiation reaction, as was done for instance in Ref.~\cite{Nagar:2019wrt}.

\acknowledgements
A. A. thanks Rossella Gamba for useful discussions on the horizon flux. A.A., V.S. and G. L.-G. have been supported by the fellowship Lumina Quaeruntur No. LQ100032102 of the Czech Academy of Sciences. A.A. is also supported by the GAUK project No. 107324. VS thanks the Charles U. \textit{Primus} Research Program 23/SCI/017 for support.

\appendix

\section{Post-Newtonian horizon fluxes}
\label{sec:PN_HF}
\begin{figure*}
\includegraphics[width=0.45\textwidth]{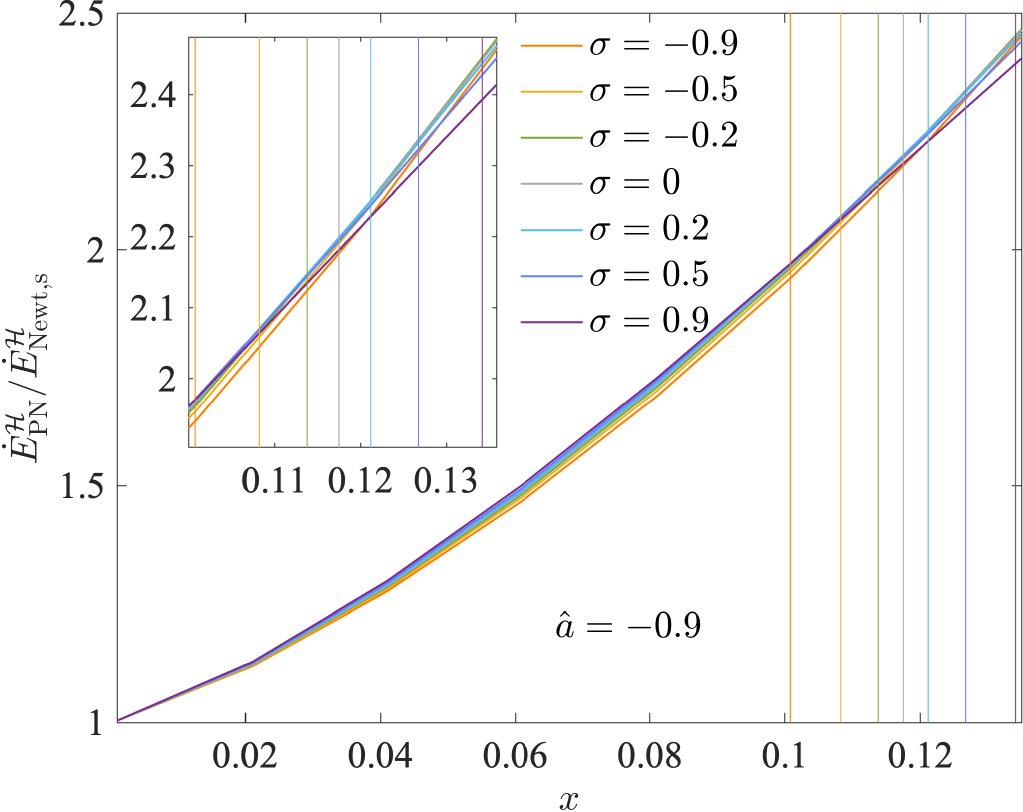} 
\includegraphics[width=0.45\textwidth]{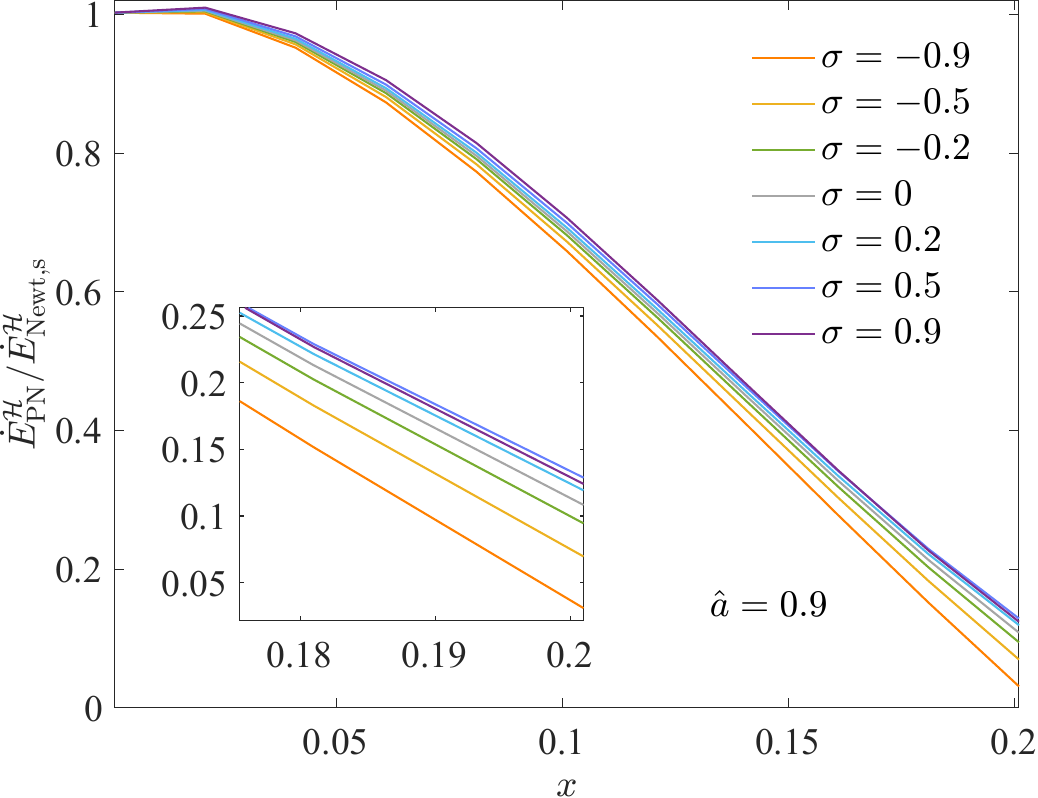} 
\caption{\label{fig:PNfluxes} Gravitational wave energy fluxes into the horizon for $\hat{a} = \{-0.9, 0.9\}$, evaluated with Eq.~\eqref{eq:PN_HF} and normaliszd by Eq.~\eqref{eq:NewtHFspin}. For $\hat{a} = -0.9$ we plot vertical lines for the MPD LSO $x$ values, and the behavior of the curves for larger frequencies is discussed in the text. For small values of $x$, positive/negative values of $sigma$ are related to a larger/smaller horizon flux, as we see for the numerical fluxes. This suggests that spin-spin contributions dominate over spin-orbit interaction in this regime.}
\end{figure*}

\begin{figure}
\includegraphics[width=0.45\textwidth]{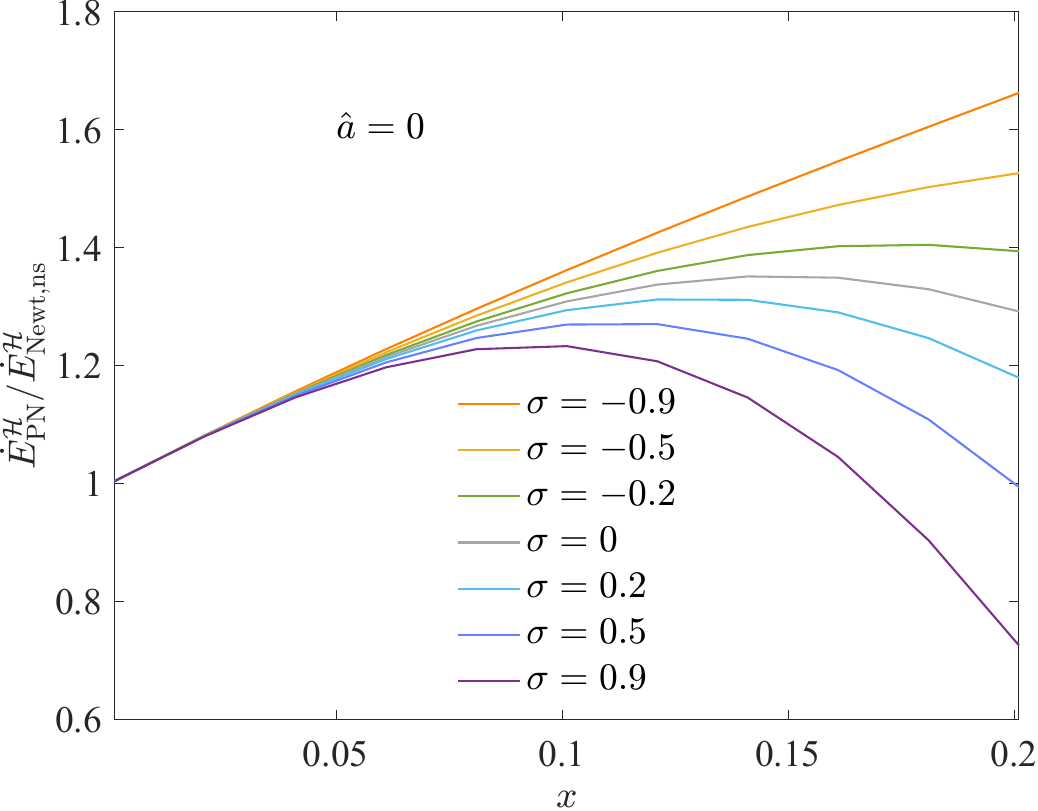} 
\caption{\label{fig:PNfluxes_a0} Gravitational wave energy fluxes into the horizon for $\hat{a} = 0$, evaluated with Eq.~\eqref{eq:PN_HF} and normalized by Eq.~\eqref{eq:NewtHFnonspin}. In the absence of spin-spin contributions, spin-orbit interaction related to the spin of the secondary yields larger/smaller fluxes for negative/positive values of $\sigma$ with respect to the nonspinning particle result.}
\end{figure}

In the main part of this work we have seen how positive/negative values $\sigma$ are related to larger/smaller horizon flux values with respect to the nonspinning particle case, which is the opposite of what we see for asymptotic fluxes. To further benchmark the behavior of the horizon flux and its dependence on $\sigma$, we take into account analytical results from PN theory. In particular, in this section we exploit the result of Ref.~\cite{Saketh:2022xjb} that has been derived for comparable-mass binaries and compute its spinning-particle limit. We note that the authors already provide a test-mass expression, that however does not incorporate the spin of the secondary. We start from their Eq.~(4.22), that reads
\be
\label{eq:PN_HF}
\left\langle \frac{dm_1}{dt} \right\rangle = \Omega(\Omega_{\rm H} - \Omega) C_x ,
\ee
where $\Omega$ is \textit{not} the orbital frequency as defined in the rest of our paper, but the angular velocity of the tidal field in the primary BH frame, $\Omega_{\rm H} = \chi_1 /[2 m_1 (1 + \kappa_1)]$ is the horizon angular velocity of the primary BH, and $\kappa_1 = \sqrt{1 - \chi_1^2}$. We consider $\sigma = \nu \chi_2$ and $\hat{a} = \chi_1$, where $\chi_i$ in their work are the dimensionless spins, namely $\chi_i = S_i/m_i^2$, and $m_1, m_2$ are respectively the mass of the primary and of the secondary. At leading order in the mass ratio, defining for convenience the factor $W \equiv 6 + 3x + (3\sigma + 2 \hat{a})x^{3/2}$, we find
\be
\Omega = \frac{x^{3/2}}{216 m_1} W^3 \left( 1- \frac{x}{24} W^2  - \frac{x^{3/2}}{432} (3\sigma + 2 \hat{a}) W^3 \right) 
\ee
and
\begin{widetext}
\be
C_x = - \frac{16}{5} x^6 (1 + \kappa_1) m_2^2 \left[ 1 + 3 \hat{a}^2 + \frac{x}{4} (16 + 33 \hat{a}^2) +  \frac{x^{3/2}}{6} \left( - (104 + 120 \kappa_1) \hat{a} + (6 - 72 \kappa_1) \hat{a}^3 + 45 \sigma \hat{a}^2 - B_2(\hat{a})(48 + 144 \hat{a}^2)\right)\right]
\ee
\end{widetext}
where $B_2 = \Im[\rm{PolyGamma}$$(0, 3+ 2i\hat{a}/\kappa_1)]$. We note the masses $m_1, m_2$ can be factorized out of the expression, so to have $m_2^2/m_1^2 \sim \nu^2$. We keep these expressions as they are, not linearizing in the secondary spin. We plot the curves evaluated from Eq.~\eqref{eq:PN_HF} and normalized by the Newtonian contribution \eqref{eq:NewtHFspin} in Fig.~\ref{fig:PNfluxes}, where we can see that at low frequencies the dependence of the horizon flux on the secondary spin is qualitatively consistent with our result, namely positive/negative values of $\sigma$ correspond to values of the flux that are larger/smaller than the non-spinning secondary result. As the orbits get closer to the LSO however, we see that for $\hat{a} = -0.9$ there is a change in this behavior. In the bottom panel of Fig.~\ref{fig:fluxes_H} we see the same thing happening for $\sigma = 0.9$, which suggests that the spin-orbit interaction related to the secondary spin prevails over the spin-spin contribution. In Fig.~\ref{fig:PNfluxes_a0} we plot the Newtonian-normalized PN flux for a nonspinning primary, and we see that in the absence of spin-spin contributions, spin-orbit interaction related to the spin of the secondary yields larger/smaller fluxes for negative/positive values of $\sigma$ with respect to the nonspinning particle result. A more detailed study of the interplay between spin-spin and spin-orbit effects in the horizon flux for frequencies closer to the LSO could be pursued by evaluating the numerical fluxes without linearizing the input dynamics in $\sigma$. We leave this analysis to future work.
 
\section{Fluxes for a spinning particle on Schwarzschild}
\label{sec:Schwarzschild} 
In this appendix we summarize the procedure followed by Ref.~\cite{Harms:2016ctx} to obtain asymptotic GW fluxes for a spinning particle on CEOs around a Schwarzschild BH, in order to point out the differences with the procedure we use in this work. 

First, the authors chose a set of BL radii $r_{\rm MPD} = \{4, 5, 6, 8, 10, 12, 20, 30\}$ and solved numerically the relation $u_{\rm EOB} = u_{\rm MPD}$,
\be
u_{\rm MPD} - \sigma u_{\rm MPD}^{5/2} = u_{\rm EOB} - \sigma \frac{u_{\rm EOB}^{5/2}}{1 + \sqrt{\frac{1-2u_{\rm EOB}}{1-3u_{\rm EOB}}}}, 
\ee
with a MATLAB routine. Then, for the given values of $u_{\rm EOB}$ the momentum $p_\varphi$ for circular orbits is evaluated by solving numerically 
\be
\label{eq:dHdu}
\frac{\p \hat{H}_{\rm eff}}{\p u_{\rm EOB}} = 0 .
\ee
The values of $u_{\rm EOB}$ and $p_\varphi$ found numerically are finally used to evaluate the energy $\hat{H}_{\rm eff}$. All these three steps are not strictly linear in $\sigma$: the values of $u_{\rm EOB}$ contain non-linear-in-$\sigma$ contributions, which are then propagated when solving Eq.~\eqref{eq:dHdu} for $p_\varphi$, and finally into $\hat{H}_{\rm eff}$. As for MPD, the values of the energy and angular momentum for CEOs (for the TD SSC) are also found numerically as the extrema of an effective potential, instead of using the analytical, linear-in-$\sigma$ expressions. This yielded a difference in the evaluated fluxes. In table \ref{tab:energy} we list the values of the energy for CEOs for $\sigma = 0.5$ at fixed BL radii $r_{\rm MPD} = \{4, 5, 6, 8, 10, 12, 20, 30\}$ for (i) EOB, following the numerical procedure of Ref.~\cite{Harms:2016ctx}; (ii) EOB, finding the energy as a function of the BL radius exploiting
\begin{align}
    x &= u_{\rm MPD} - \sigma u_{\rm MPD}^{5/2}, \\
    \hat{E}_{\rm circ}(x) &= \frac{1-2x}{\sqrt{1-3x}} - \sigma \frac{x^{5/2}}{\sqrt{1-3x}},
\end{align}
where the second one corresponds to Eq.~(82) in Ref.~\cite{Harms:2016ctx}, and we linearize it as
\begin{align}
\label{eq:HEOBlin}
\hat{E}_{\rm circ, lin}(u_{\rm MPD}) &= \frac{1-2u_{\rm MPD}}{\sqrt{1-3u_{\rm MPD}}} \nonumber \\
&- \sigma \frac{u_{\rm MPD}^{5/2}}{2 (1-3u_{\rm MPD}) \sqrt{1-3u_{\rm MPD}}} ;
\end{align}
(iii) MPD, using the full, non-linear-in-$\sigma$ analytical expression~\cite{Hackmann:2014tga}; (iv) MPD, using the analytical expression linearized in $\sigma$. The reader can check that the linearized-in-$\sigma$ expressions agree, which explains why working consistently with the linearized quantities led us to find the same evaluated fluxes for EOB and MPD when the central BH is nonspinning. 

\begin{table*}[htp!]
\begin{center}
\begin{ruledtabular}
\begin{tabular}{c | c c | c c}
$r_{\rm MPD}$ & $\hat{H}_{\rm eff}^{\rm num}$ & $\hat{E}_{\rm circ, lin}$ & $E_{\rm MPD}^{\rm full}$ & $E_{\rm MPD}^{\rm lin}$  \\
\hline 
\hline
4 & 0.9506861201 & 0.9375000000 & 0.9465872266 & 0.9375000000 \\
5 & 0.9326275668 & 0.9310056285 & 0.9319288795 & 0.9310056285 \\
6 & 0.9348732586 & 0.9347902878 & 0.9349517612 & 0.9347902878 \\
8 & 0.9455984811 & 0.9458882131 & 0.9458868418 & 0.9458882131 \\
10 & 0.9545911149 & 0.9548330143 & 0.9548254522 & 0.9548330143 \\
12 & 0.9612997024 & 0.9614788437 & 0.9614736064 & 0.9614788437 \\
20 & 0.9759476947 & 0.9760087252 & 0.9760077670 & 0.9760087252 \\
30 & 0.9837366968 & 0.9837603183 & 0.9837601107 & 0.9837603183 \\
\end{tabular}
\end{ruledtabular}
\end{center} 
\caption{Values of the energy for CEOs for $\sigma = 0.5$ for the values of the BL radius given in the first column. The second and third columns are respectively the EOB energy found with the numerical procedure of Ref.~\cite{Harms:2016ctx} and the EOB energy found with the linearized-in-$\sigma$ analytical expression, Eq.~\eqref{eq:HEOBlin}. The fourth and fifth columns contain values for the MPD energy found respectively with the full non-linear-in-$\sigma$ analytical expression~\cite{Hackmann:2014tga} and with the linear-in-$\sigma$ one. The reader can verify that the EOB and MPD linearized expressions yield the same values of the energy.}
\label{tab:energy}
\end{table*}

\bibliography{refs20241220.bib,local.bib}

\end{document}